\documentclass[manuscript,authorversion,nonacm,acmsmall]{acmart}%
\usepackage{soul}%
\usepackage[normalem]{ulem}%
\usepackage{amsmath}
\usepackage[export]{adjustbox}%
\usepackage[inline]{enumitem}%
\usepackage{rotating}%
\usepackage{tabularx}%
\usepackage{framed}%
\usepackage{makecell}%
\usepackage{pifont}
\newcommand{\circnum}[1]{\ding{\numexpr 171 + #1\relax}}
\usepackage{color}%
\newcommand{\todo}[1]{{\leavevmode\color{black}#1}}%
\newcommand{\ok}[1]{{\leavevmode\color{black}#1}}%
\newcommand{\transition}[1]{{\leavevmode\color{black}#1}}%
\usepackage{comment}%
\usepackage{caption}%
\usepackage{subcaption}%
\usepackage{tikz}%
\usetikzlibrary{shapes.geometric, arrows}%
\usepackage{etoolbox}%
\newtoggle{anonymous}
\makeatletter
\@ifclasswith{acmart}{anonymous}{
    \settoggle{anonymous}{true}
}{
    \settoggle{anonymous}{false}
}
\makeatother
\settoggle{anonymous}{false}
\setcopyright{none}%
\authorsaddresses{}%
\setcopyright{rightsretained}%
\acmYear{2025}%
\copyrightyear{2025}%
\acmDOI{XXXXXXX.XXXXXXX}
\acmJournal{JACM}
\acmVolume{37}
\acmNumber{4}
\acmArticle{111}
\acmMonth{8}

\begin{document}

\title[Artworks Reimagined]{Artworks Reimagined:
    Exploring Human-AI Co-Creation through Body Prompting%
}

\author{Jonas Oppenlaender}
\email{jonas.oppenlaender@oulu.fi}
\orcid{0000-0002-2342-1540}
\affiliation{%
  \institution{University of Oulu}
  \city{Oulu}
  \country{Finland}
}

\author{Hannah Johnston}
\email{hannahjohnston@cmail.carleton.ca}
\affiliation{%
  \institution{Carleton University}
  \city{Ottawa}
  \country{Canada}
}

\author{Johanna Silvennoinen}
\email{johanna.silvennoinen@jyu.fi}
\orcid{0000-0002-0763-0297}
\affiliation{%
  \institution{University of Jyv\"askyl\"a}
  \city{Jyv\"askyl\"a}
  \country{Finland}
}

\author{Helena Barranha}
\email{helenabarranha@tecnico.ulisboa.pt}
\orcid{0000-0003-0250-1020}
\affiliation{%
  \institution{IST, University of Lisbon \& 
    IHA-NOVA FCSH / IN2PAST
  }
  \city{Lisbon}
  \country{Portugal}
}


\begin{abstract}
Image generation using generative artificial intelligence has become a popular activity.
However, text-to-image generation---where images are produced from typed prompts---can be less engaging in public settings since the act of typing tends to limit interactive audience participation, thereby reducing its suitability for designing dynamic public installations.
In this article, we explore \textit{body prompting} as input modality for image generation in the context of installations at public event settings.
    Body prompting extends interaction with generative AI beyond textual inputs to reconnect the creative act of image generation with the physical act of creating artworks.
We implement this concept in an interactive art installation, \textit{Artworks Reimagined}, designed to transform existing artworks via body prompting.
We deployed 
the installation at an 
event with hundreds of visitors in a public and private setting.
Our semi-structured interviews with a sample of visitors ($N=79$) show that body prompting was well-received and provides an engaging and fun experience to the installation's visitors.
We present insights into participants' experience of body prompting and AI co-creation
and
identify three distinct strategies of embodied interaction
focused on re-creating, reimagining, or casual interaction.
We provide valuable recommendations for practitioners 
seeking to design interactive generative AI experiences 
in museums, galleries, and public event spaces.
\end{abstract}



\maketitle

\section{Introduction}%
\label{sec:introduction}%

Recent advances in generative artificial intelligence (AI) have fueled widespread interest in text-to-image generation systems, which millions of users now employ generative AI for creative exploration and recreation \cite{creativity,wired,palgrave,3591196.3593051.pdf}. 
Yet these systems are typically used in private settings where users type discrete prompts on keyboards, a design that limits engagement in public contexts.
This reliance on text-based input brings four key challenges—input control, literacy, social creativity, and physical setting constraints (detailed in Section 3.2)—that present barriers to accessibility and limit the potential for interactive, public experiences~\cite{3464327.3464965.pdf}.

Building on earlier efforts to involve audiences in image generation in public settings, such as GenFrame \cite{Kun2024} in a museum setting, researchers have begun exploring alternative input methods for image generation.
    GenFrame, for instance, allows visitors to influence the image creation process through three rotating knobs, offering a novel interaction modality to image generation \cite{Kun2024,3643834.3660750.pdf}.
However, this approach 
    restricts human agency because much of the creative decision-making remains with the AI system rather than with the user.
Our work aims to address these limitations by exploring an interaction design that more effectively balances user input with algorithmic creativity, ultimately making image generation more inclusive and engaging in public events.

In this paper, we introduce \textbf{body prompting} as a novel input modality for image generation. Instead of relying solely on text, body prompting uses a person’s pose to guide the generative process, reconnecting image creation with the expressive physicality of traditional art-making.
Our work, thus, touches on 
considerations of embodiment, materiality, and material agency that have long informed the creation of traditional visual art \citep{978-3-031-29956-8_13.pdf}.

Our goal is to empower audiences in public settings to ``reimagine'' existing artworks through their body movements, creating personalized reinterpretations.
To support this vision, we designed and implemented an interactive art installation—\textbf{Artworks Reimagined}—which leverages generative AI to transform classical artworks based on users’ body prompts. The installation was deployed at the European Researchers’ Night 2023, a public event that attracts thousands of visitors.

We evaluated body prompting in a field study at European Researchers’ Night, engaging hundreds of visitors. Acknowledging that some individuals may feel inhibited when performing in front of others \cite{quiet}, we implemented two interaction settings: a public staging area for body prompting in direct view of an audience, and a private booth where visitors could interact and view results without an on-looking audience (cf. \autoref{fig:setup}).
Our investigation combines observations of participants’ interactions and strategies with follow-up semi-structured interviews ($N=79$), offering insights into how visitors naturally engage with generative technology via body prompting in a public context without structured guidance.
%
%
Our 
evaluation of the system is guided by the following research questions:%
\begin{itemize}%
    \item[RQ1:] \textit{Generated images:}
        What images do visitors create with body prompting? Is there a difference in images between the public and private setting?
    \item[RQ2:] \textit{User experience:}
        How do visitors experience body prompting?
        And how do they experience AI co-creation via body prompting?
    \item[RQ3:] \textit{Participant behavior:}
        What are the decisions and preferences of visitors in regard to body prompting? 
        What goals do visitors pursue when body prompting?
    \item[RQ4:] \textit{Body prompting in public:}
        How do personality traits, as measured by the ``Big Five'' \citep{BigFive}, affect body prompting and the use of the art installation in the private and public setting?
\end{itemize}%




\ok{%
Our exploration of body prompting as an interaction method for generative AI art installations at public events indicates that this approach effectively engages both active participants and observers.
Body prompting was perceived as an engaging and fun experience.
Participants employed three distinct strategies: imitating the source artwork, reimagining it, and expressing themselves naturally.
Notably, personality traits did not significantly influence the choice of posing strategy or setting, as both extroverts and introverts appreciated the experience.
}%

We contribute to the field of Engineering Interactive Computing Systems (EICS) by engineering a novel interactive system that uses body prompting to drive generative AI.
Our work describes a modular pipeline and software architecture based on T2I-Adapter with Stable Diffusion---combining ControlNet-based pose detection, style transfer, and CLIP-Interrogator for automated prompt generation---with dual web applications and cloud services (AWS Gateway, Lambda, S3) to generate personalized recreations and reimaginations of an existing digital artwork. 
This work advances engineering methods by providing a detailed architecture, workflow processes, and evaluation of user strategies that can be adapted to design and deploy co-creative interactive systems with generative AI in public settings.
In addition, we make design recommendations for developing and refining 
interactive computing systems in dynamic, real-world settings.

The remainder of this work is structured as follows.
After reviewing related work in Section \ref{sec:relatedwork},
we present the design rationale motivating our design of Artworks Reimagined in Section \ref{sec:method}.
We present the installation's overall concept, design, and implementation in Section \ref{sec:system}.
We then describe our evaluation of the system in Section \ref{sec:procedure}. 
this includes an overview of the images generated, detailed analysis of interview responses,
and a synthesis of observations from the event. 
We discuss our findings and provide design recommendations for practitioners and researchers in Section \ref{sec:discussion} and conclude in Section \ref{sec:conclusion}.

\section{Related Work}%
\label{sec:relatedwork}%

\ok{%
Image generation is typically confined to private spaces where users type prompts, consisting of discrete words, on a keyboard. This results in an isolated, individual experience \citep{3613905.3650947.pdf}.
Recent research, however, has explored alternative input modalities and interaction paradigms to integrate image generation into public and social contexts.
In this section, we review work that challenges the traditional text-based approach to image generation, and we review related work that presents challenges to its implementation.
We first discuss challenges around interactive public displays (Section \ref{sec:public-displays}), then examine embodied experiences (Section \ref{sec:interactive-art-experience}), followed by a review of studies on re-creating and reinterpreting artworks (Section \ref{sec:recreating-art}). Finally, we review selected studies 
that leveraged generative AI in public installations (Section \ref{sec:public-genai}).
}%

\subsection{Interaction with Public Displays}%
\label{sec:public-displays}%
\ok{%
Public displays offer opportunities to study how users interact with technology. However, public displays bring unique challenges, such as display blindness \citep{Pervasive09MuellerDisplayBlindness.pdf}, interaction blindness \citep{ieeepc16.pdf}, and 
the first-click problem \citep{kukka2013.pdf}.
These issues arise when users are unsure how to initiate interaction with a public display, or when they ignore the interactivity of the display entirely.
In this section, we focus on challenges specifically related to the public setting of our study.

%
Privacy concerns require consideration in the design of public display installations.
\citet{p1-brudy.pdf} identified `shoulder surfing'---where unauthorized individuals may view sensitive information---as a pertinent risk associated with public displays. 
Further exploring privacy issues, \citet{p171-Memarovic.pdf} examined how the public's concerns about photography in the vicinity of public displays influence user interaction. 
Their study emphasized the importance of transparent policies regarding the storage and usage of photos taken in these contexts, highlighting a general need for clear communication about the handling of users' data.
In a comparative study, \citet{10-1108_JSM-04-2012-0071.pdf} addressed differences between private and public self-service technologies. 
Their findings suggest that perceptions of control and convenience significantly vary depending on the setting, with public technologies often viewed as less controllable and convenient than their private counterparts \citep{10-1108_JSM-04-2012-0071.pdf}. This perception could influence user engagement and satisfaction with the public display in our study.
\citeauthor{3321335.3324943.pdf} studied
privacy in
interactive public displays \citep{3321335.3324943.pdf,Alorwu2022}.
They found that people are surprisingly
willing to volunteer
highly sensitive personal information in public settings.

These studies 
reveal 
the complexity of user behavior and expectations
when interacting with displays in public spaces.
Addressing these concerns
is crucial for designing  
an interactive public installation.%
%
}
%
\subsection{Interactive Experiences via Embodied Interaction}%
\label{sec:interactive-art-experience}%
\ok{%
Embodied interaction 
can provide an engaging experience.
This is particularly relevant in the context of public cultural spaces, such as 
GLAM institutions (Galleries, Libraries, Archives, and Museums).
In museums, 
the contact with more traditional art forms, namely painting or sculpture, tends to 
occur passively through a cognitive engagement with the artworks. 
Digitally enriched museum experiences \citep{09647775.2023.2235683.pdf}
open new opportunities to experience and interact with such artworks through embodied cognitive engagement 
via
human-technology interaction.
Gestures and posing are an engaging approach to meaning-making in the GLAM context \citep{1-s2.0-S2210656115000331-main.pdf,21-2-steier.pdf,3643834.3660711.pdf}.
In the remainder of this section, we review prior work on embodied interaction in the context of GLAM institutions and other public settings.

\citet{3313831.3376186.pdf} explored engaging museum visitors with embodied interaction. 
Their study was aimed at human-data interaction, enabling visitors to explore data sets and interactive visualizations with gestures and body movements.
\citet{duplex} presented a system for museum visitors to provide interactive feedback for artworks, including taking selfie photos and selfie videos. 
The latter was perceived as a surprising way of interacting with an artwork and its artist.
\citet{fpsyg-08-00731.pdf} also explored museum selfies, finding selfies to be a way to construct narratives about oneself.
However, 
their study
did not feature embodied interaction with dynamically generated artifacts in the museum context.


Positive effects of interactive technological engagement with artworks have been reported. 
For example, aesthetic appreciation (including interest, intensity, pleasure, and learning) was higher in artworks with interactive elements compared to non-interactive physical objects \citep{jonauskaite2024interactive}.
%
%
In \citeauthor{Coeckelbergh}'s framework \cite{Coeckelbergh}, body prompting can be conceptualized as
a performance-based approach to interaction and AI image generation.
Body prompting is a poietic performance within human-technology co-creation. The performance-based approach emphasizes the process of interaction in which novel artistic (quasi-)subjects and objects emerge and are produced in and by the processes instead of instruments and tools~\cite{Coeckelbergh}.%
}
%
\subsection{Re-creating and Reinterpreting Artworks}%
\label{sec:recreating-art}%
\ok{%
Art museums are increasingly digitizing their collections and making them accessible on the internet
thus fostering
different reinterpretations and re-creations of the paintings
\citep{Barranha-2018}.
Several experiments have shown that people enjoy re-creating and reinterpreting famous artworks.
Launched at the height of the COVID-19 pandemic, The Getty Museum Challenge \cite{getty-challenge} was a widely successful initiative, which invited the public to re-create an artwork with household items.
%
In 2023, the Mauritshuis Museum in The Hague launched another challenge, which invited the public to reinterpret Vermeer's famous painting ``The Girl with a Pearl Earring'' using different processes, including generative AI \citep{mygirlwithpearl}.
The AI-generated images were later exhibited in the Museum in place of Vermeer's famous painting,  in a digital loop display which featured a wide range of experiments produced with different media.
In parallel to such institutional initiatives, 
generative AI has also been used by researchers to reinterpret artists’ self-portraits~\citep{Barranha-2023}.

These examples of re-creating and reinterpreting artworks suggest that new activities designed to explore art collections with the use of generative AI can foster participatory creative experiences.
However, the Getty and Mauritshuis Museum challenges were completed from the participants' homes, re-creating or reinterpreting the artwork either with body poses and household items or generative AI.
In our work, we explore how visitors can be engaged to re-create or reinterpret an artwork with their bodies and generative AI at a public event.%
}%
%
\subsection{Generative AI in Public Exhibitions}%
\label{sec:public-genai}%
\ok{%
Generative AI has been 
explored in the context of exhibitions, GLAM settings, and other public spaces.
In this section, we review three selected interactive installations that used generative AI for engaging audiences.
%
%
%

The artist duo Varvara \& Mar
presented an art installation in which the human impact on a 
landscape is gradually revealed using eye-tracking and generative AI \citep{3623509.3635319}.
While their interactive art installation requires bodily presence, it is driven by 
gaze-based interaction with the public display, not body posing.

\citeauthor{3623509.3635318} explored embodiment in Shadowplay \cite{3623509.3635318}.
Users of this AI art installation produced shadows on a wall, which served as input for the image generation AI.
Our system differs in that the users' body directly serves as input 
to re-create or reinterpret an existing artwork via body prompting.



GenFrame by \citeauthor{Kun2024} is a recent related study featuring a
generative art piece designed for the museum context \citep{Kun2024,3643834.3660750.pdf}.
GenFrame's framed 
display gives the appearance of a traditional artwork, however the audience can control the image generation with three rotating knobs.
Like our work, GenFrame was implemented with ControlNet and StableDiffusion.
Nevertheless, the interaction affordances of GenFrame are limited by design. Our work explores body prompting as an expressive way of providing input to the image generation process.%
%
%
%
}%
%
%
\section{System Design}%
\label{sec:method}%
\todo{%
In this section, we first provide background information on the public event that was the context for our study and our motivation and design goals for Artworks Reimagined in Section \ref{sec:researchers-night}.
We then describe four key challenges based on our formative study in Section \ref{sec:limits} and the design goals of our system in Section \ref{sec:designgoals}.
}%
%
\subsection{Formative Study}%
\label{sec:researchers-night}%
\ok{%
Our work was informed by our
prior experience of hosting an exhibit at the European Researchers' Night in 2022.
The European Researchers' Night is an initiative funded by the European Union under the Marie Skłodowska-Curie Actions (MSCA).
It is the largest science outreach event in Europe, aiming to bring researchers closer to the public and showcasing the impact of research on daily life. 
The event takes place annually 
at universities in numerous European countries. 
The European Researchers' Night
aims to
    interest citizens 
    in research.
To this end, the event gives the public the opportunity to discover the wonders of science in fun and inspiring ways. 
As a multidisciplinary event, Researchers’ Night offers a wide range of activities from workshops to laboratories, presentations, lectures, and exhibits.
The event program caters to the general public, including visiting school groups and parents with young children.
}%

\ok{%
Text-to-image generation was still a novel phenomenon in 2022. 
We decided to demonstrate this novel technology to visitors of the European Researchers' Night.
Our exhibition stand at the event 
allowed visitors to try out three existing text-to-image generation systems (Midjourney, Stable Diffusion, and DALL-E).
The exhibition stand consisted of two cocktail tables with one laptop each.
One of the laptops was connected to a large public display, the other allowed image generation more privately without displaying results publically. 
}%

\ok{%
During this well-visited event,
we had numerous conversations with visitors, noting that many visitors had not tried text-to-image generation before.
One central insight from our observations and conversations with visitors was that,
while interesting for the person writing the prompts, text-based interaction with the image generation system was not very appealing and engaging for an audience in a public setting.
While prompts were being written in private, and users were deeply engaged in writing these prompts, other visitors could simply not see what was going on. 
The interaction with the generative AI was hidden from the public eye.
For passers-by, the mechanics of the public display installation were not obvious.
That is, it was not immediately obvious that prompts written in private on the laptop manifested as images on the large screen.
}%

\ok{%
For the following year,
we built on this important observation and designed an interactive art installation to address this issue.
In particular, we sought to address the ``air gap'' between input and screen, to provide a more interactive and expressive experience to the users, and a more engaging experience for all other visitors.
\transition{In the following section, we summarize our formative study's findings by outlining four key challenges that limit the application of text-to-image generation in public event settings.}%
}%

\subsection{Challenges of Text-to-Image Generation for Public Event Settings}
\label{sec:limits}
\ok{%
From the formative study, we distill 
four key
challenges of text-based image generation that limit its use in public event settings:
    output control,
    literacy,
    collaboration \& social creativity,
    and
    expressivity. 
\begin{itemize}
    \item[C1.] 
    \textit{Output control:}
            The level of control over text-to-image generation is limited by language~\citep{978-3-031-29956-8_13.pdf}.
            One cannot describe an image with discrete prompts in every detail, leaving much of the initial image generation to randomness \citep{Spectrum.pdf}.
            ``Prompt engineering'' \cite{palgrave,creativity,modifiers,oppenlaender2023prompting} for text-to-image generation  is tedious, and controlling the image generation process requires a considerable amount of practice and skill \cite{oppenlaender2023prompting}.


    \item[C2.] 
    \textit{Literacy:}
            Frequently regarded as an instrumentalist use of technology,
            text-to-image generation is a process in which AI is used as a tool to visually translate concepts or ideas \citep{Coeckelbergh}.
            For this translation to
            be effective, English literacy is required.
            Further, some image generators require the use of specific  keywords to produce high-quality outputs \citep{modifiers},  which demands expertise in ``prompt engineering'' 
            and familiarity with art concepts, such as  
            styles and media \citep{modifiers}.
            These language-based requirements
            create barriers of accessibility
            and
            exclude certain populations---such as the illiterate, young children, and all those who are not familiar with the English language---from participating.

    \item[C3.] 
    \textit{Collaboration \& social creativity:}
            Contemporary literature views creativity as a social activity \citep{2019_chi-paper.pdf,amabile}, with the concept of the lone creator being dispelled as a myth \citep{10.2307.23216799.pdf}.
            Yet, image production with generative AI is almost exclusively completed at home by a solitary creator.
            The social and performative aspects of art creation are lost due to the privacy of the image generation setting. 
            The text input modality is not very engaging for an audience in a public event setting.
            Collaborative websites, such as Artbreeder\footnote{https://www.artbreeder.com},
            have demonstrated that collaborative image generation can be an engaging experience, with wide audience appeal.%

    \item[C4.] 
    \textit{Expressivity:}
            Confined to the 
            narrow triangular interaction space between user, keyboard, and computer display,
            the practice of text-to-image generation does not lend itself well to public settings, such as
            Galleries, Libraries, Archives, and Museums (GLAM).
            For such public settings, input modalities 
            should be dynamic and intuitive, offering glimpses of the generative process to the audience and building anticipation for the final image.
            For a public event setting, engaging larger groups of people beyond the individual, is required.
            Typing on a keyboard in front of a computer screen is not expressive, in this regard, and
            does not provide an engaging experience to an audience.

\end{itemize}

These challenges, while not being insurmountable in the use of text-to-image generation by an individual practitioner from home,
present challenges to the use of text-to-image generation technology in 
installations at public event settings.  
\transition{%
In the following sections, we explain the design goals for our interactive art installation (Section \ref{sec:designgoals}),
and how we aimed to address these challenges in the design of our installation (Section \ref{sec:system}). 
}%
}

\subsection{Design Goals}
\label{sec:designgoals}

\ok{%
In our design of the interactive art installation, we sought to overcome the limitations in engagement that
we had observed at the prior European Researchers' Night event.
Our aim was to provide a highly engaging experience for visitors of this event.
Based on our formative study
and the four challenges delineated above,
we formulate the following five design goals for our art installation in a public event setting:

\begin{itemize}%
    \item[D1]
    \textit{Enhance user control over image generation:}
    By allowing body movements to guide image creation, the installation gives users a direct and intuitive way to shape outcomes. This design goal addresses the challenge of output control (C1) by reducing reliance on imprecise text prompts and enabling more immediate and expressive creative decisions (C4).

    \item[D2]
    \textit{Reduce language dependency:}
    Leveraging body prompting removes the need for specialized textual skills or prompt engineering (C2).
    This makes the experience more inclusive by lowering language and literacy barriers, ensuring that people of different ages, language proficiencies, and technical backgrounds can participate.
    
    \item[D3]
    \textit{Facilitate social creativity and collaboration:}
    The system is designed to encourage group participation and shared creative exploration (C3).
    By shifting from solitary, text-based input to embodied interaction, it supports collaborative artistic expression and reintroduces the social dimension of art-making, which is often lost in 
    generative AI setups.
    
    \item[D4]
    \textit{Support dynamic and expressive interactions:}
    Public installations benefit from interactions that are visibly engaging and performative. This goal ensures that the system’s real-time feedback and visual display 
    supports the posing process and
    creates an engaging experience for both active participants and observers, addressing the expressivity challenge (C4) inherent in static text-based interfaces.
    
    \item[D5]
    \textit{Accommodate varied interaction contexts:}
    Recognizing that some users may feel inhibited in public performance, the installation provides both public staging and private booth settings. This design goal aims to offer flexibility so that users can choose the interaction context that best suits their comfort level, thereby fostering participation across a diverse audience.
\end{itemize}%
    
%
%

The design goals outlined above provided a 
framework for addressing the challenges of text-based image generation in public installations.
\transition{%
We now turn 
to the system design and implementation, detailing how our architecture and technical choices align with the design goals.
In the following section, we describe the system's overall concept, its modular pipeline and key components, and interaction mechanisms that enable embodied co-creation in dynamic public settings.
}%
}

\section{Artworks Reimagined}%
\label{sec:system}%
We designed and developed an interactive art installation, called \textit{Artworks Reimagined}, that allows its users to engage with image generation technology with a low barrier of entry.


\subsection{Overall Concept Addressing the Design Goals}%
\label{sec:OverallConcept}%
\label{sec:designgoalssolutions}%

Artworks Reimagined is an interactive installation that leverages ``body prompting'' to enable users to reimagine classical artworks through their own physical expressions.
Rather than relying on traditional text-based inputs, visitors use their body poses to guide a generative AI, resulting in 
transformations of existing digital artwork images.
This approach reconnects the creative process with embodied art-making, offering a more natural and expressive means for users to interact with technology.
The installation was designed for public events, featuring both an open staging area and a private booth to accommodate different comfort levels.
By integrating 
pose detection with a generative pipeline, Artworks Reimagined invites visitors to engage in a collaborative and dynamic reinterpretation of art, bridging the gap between traditional creative practices and contemporary AI technologies.

\subsection{Materials}%
\label{sec:sourceartworks}%
\ok{%
As source materials for the interactive art installation, we aimed to select materials that would be interesting and engaging to visitors.
To this end, we selected digital artworks from two sources.

The first source included paintings from the \textit{Golden Age of Finnish Art} (%
circa 1880-1910).
    The Golden Age of Finnish Art is central to Finnish identity, as it represents a transformative era when artists celebrated Finland’s unique natural beauty, cultural heritage, and emerging national spirit, thereby reinforcing a strong sense of pride and self-determination.
We identified artworks from this era by visiting Wikipedia pages of prominent artists from that time period.\footnote{See https://en.wikipedia.org/wiki/Golden\_Age\_of\_Finnish\_Art}

The second source was a list of popular paintings on \textit{WikiArt}.\footnote{See https://www.wikiart.org/en/popular-paintings/alltime}
We purposefully selected artworks from this list, taking care to avoid including nudity and other depictions not appropriate for public viewing in the presence of children.

This convenience sample of 60 artworks (29 from the Finnish Golden Age and 31 from WikiArt) represents a 
mix of well-known international masterpieces and artworks with a deep local context.
}%


\subsection{Technical Implementation}%
\begin{figure*}[htb]%
  \centering%
  \includegraphics[width=\linewidth]{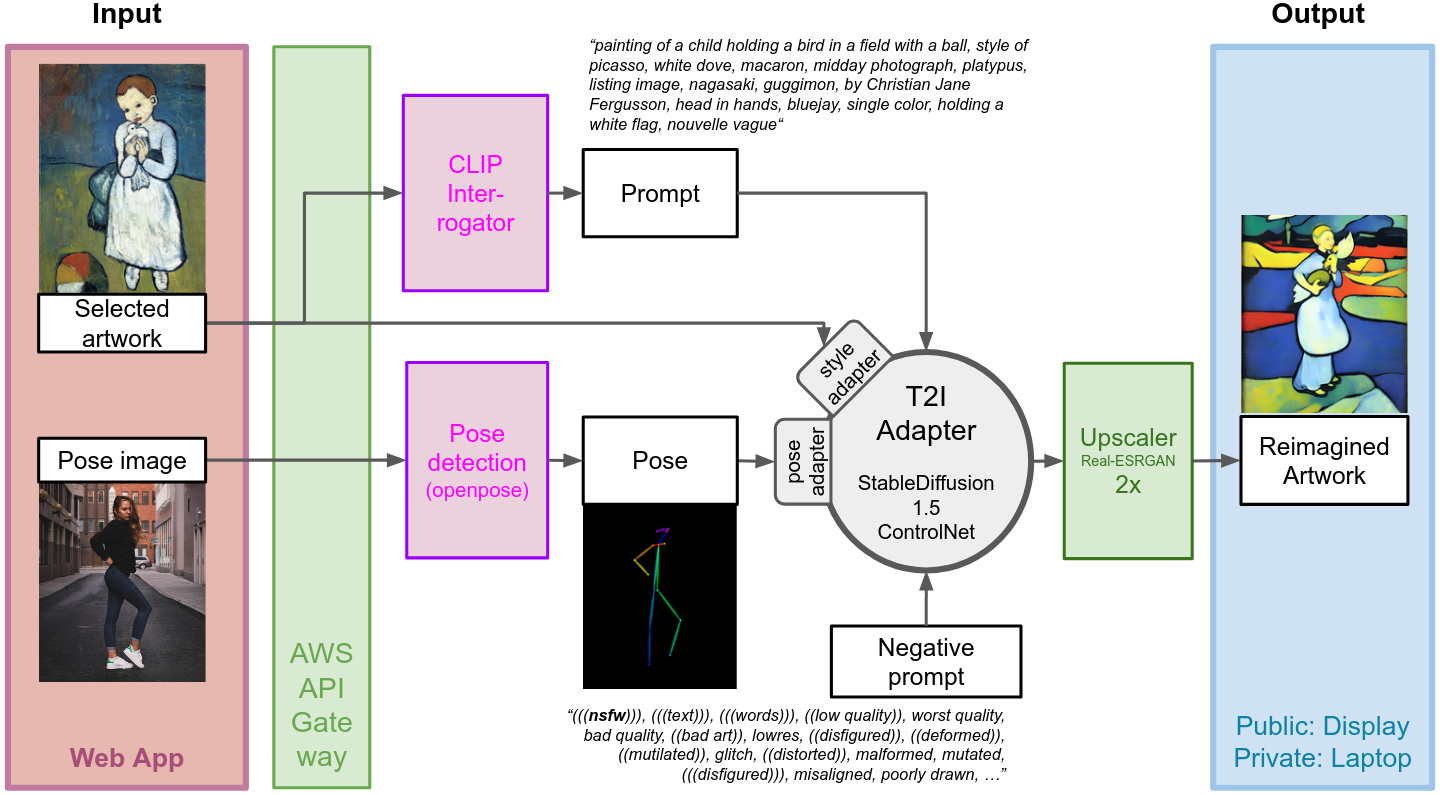}
  \caption{Application architecture with inputs (left), 
  textual prompts and body prompt (middle), and output (right).}
  \label{fig:architecture}%
\end{figure*}

\label{sec:implementation}%
\ok{%
We developed a software architecture (see \autoref{fig:architecture})
enabling us to generate images based on four inputs:
    the user's pose,
    the existing artwork,
    and two prompts.
This architecture allows us to generate a digital image that resembles the style of the original artwork, but is adapted to a different pose.
\transition{%
In Section \ref{sec:imagegen}, we provide details on the implementation of our system's backend, where images are created via API. This is followed by a description of the frontend (user interface) in Section \ref{sec:apps}.
}%
}%

\subsubsection{Output control}%
\label{sec:controlnet}
            In the context of image generation with generative AI,
            ControlNet~\citep{ControlNet} was a remarkable breakthrough, allowing fine-grained control over image generation via human poses and other input types (such as depth masks, edge maps, and line drawings).
            A pose in ControlNet is a skeleton-like structure 
            with key points that correspond to major joints and body parts.
Our installation uses ControlNet to translate captured body poses into structured constraints that guide the image synthesis process. Specifically, a real-time camera feed allows users to assume a specific pose.
A photo of the user's pose is taken from this stream, and a pose detection algorithm (OpenPose~\cite{openpose}) extracts key points to form a skeletal representation of the user's body prompt.
This representation is then provided as input to ControlNet, which generates an image layout aligned with the user's pose.
the generative processing also  integrates style transfer techniques and automated prompt refinement to balance user control with algorithmic creativity.
This layered approach ensures that the final output reflects both the expressive intent of the user's movements and the inherent unpredictability of generative AI, thereby addressing the challenge of output control in public art installations (C1).

\subsubsection{Backend for image generation}%
\label{sec:imagegen}
\ok{%
The central component in our architecture is~T2I-Adapter~\citep{2302.08453.pdf}.
    T2I-Adapter is a controllable image generation framework that extends text-to-image (T2I) diffusion models by incorporating additional structural guidance, such as edges, depth, or keypoints, to better align generated images with user inputs.
``Adapters'' are used by T2I-adapter to receive the inputs.
In our case, T2I-adapter receives four different inputs, as follows
    (cf. \autoref{fig:architecture}):
}%
\ok{%
\begin{enumerate}
    \item \textit{Pose}: A webcam image took a photo of the user. The user's pose was then detected from this body prompt photo with openpose-based 
    ControlNet~\citep{ControlNet,openpose}.
    In our system, the pose did not include facial expression and hand movements for improved anonymity.
    The pose was then used to guide the image generation via T2I-Adapter's pose adapter.
    \item \textit{Artwork}: Style information was extracted from the 
    source artwork and used as an input for the T2I-Adapter, using the Style Adapter in T2I-Adapter.
        The Style Adapter is a module that guides text-to-image diffusion models to generate images with a specific artistic or visual style by conditioning the generation process on reference style features.
    \item \textit{Prompt}: A textual prompt was generated from the selected source artwork using CLIP-Interrogator (version 0.6.0).\footnote{https://github.com/pharmapsychotic/clip-interrogator}
    CLIP-Interrogator is a tool that uses OpenAI's CLIP \cite{CLIP} model to analyze an image and generate a descriptive text prompt that could have produced it.
    This prompt was given to T2I-Adapter to guide the image generation, in addition to the other inputs. 
    \item \textit{Negative prompt}: A negative textual prompt was manually designed, based on best practices in text-to-image generation \citep{modifiers}.
        Negative prompts are commonly used to mitigate generative glitches, prevent certain subjects from appearing, and to avoid low quality outputs.
    The same negative prompt was used for all image generations to improve the quality of outputs and prevent 
    unsafe generations~\citep{2305.13873.pdf}.%
\end{enumerate}%
}%

\ok{%
Within T2I-Adapter, images were generated with Stable Diffusion version 1.5 \citep{rombach2022highresolution} with 50~diffusion steps and CFG 8.0.
    The CFG (Classifier-Free Guidance) parameter in text-to-image generation controls the balance between faithfulness to the prompt and diversity in the generated images.
The generated images were upscaled twice with Real-ESRGAN \cite{Real-ESRGAN} to increase the images' resolution for viewing on the large public display. GFPGAN \cite{GFPGAN} was used to enhance faces.
The image generation architecture was packaged as a cog container \cite{cog} and uploaded to Replicate.com where it is publicly available 
via a user interface and an API.\footnote{%
\href{https://replicate.com/joetm/camerabooth-openpose-style}{https://replicate.com/joetm/camerabooth-openpose-style}
}
}

\subsubsection{Frontends for guiding the process and viewing the results}
\label{sec:apps}
\ok{%
Two separate web applications were developed, one for guiding the body prompting process and one for viewing the results.
We decided on two applications because we wanted to avoid users from blocking the body prompting stage when viewing results. Instead, users viewed the results on a separate display (cf. \autoref{fig:setup}).

Both applications were created with React, a JavaScript web development framework, and hosted as static websites in AWS S3 buckets, using an ssl-encrypted AWS Cloudfront to provide a secure connection.
The secure connection was required by the MediaStream Web API used for acquiring images from the webcam.
Much of our installation's functionality will become clearer in our description of the user study in Section \ref{sec:procedure}.
In the remainder of this section, we describe the technical implementation of the two applications.
}%

\subsubsection*{Web Application 1: Body prompting}
\ok{%
The first application guided users through the process
with a linear application flow (c.f. \autoref{fig:process}).
After providing consent, a timer was started on the click of a button.
Once the timer hit zero, the application captured the 
photo from the webcam MediaStream using the MediaDevices API.
The processing of the photo was then completed in the cloud, by sending an API request with the image as payload (in binary format) to Replicate's servers.
    Replicate\footnote{https://www.replicate.com} is a platform that enables users to run machine learning models in the cloud via a simple API, providing access to various AI models without requiring local setup.

An AWS Gateway was set up to proxy API requests from our web application to Replicate's API. This was done to avoid issues with CORS (cross-origin resource sharing) in our web application.
Once completed, results from Replicate's API were received with a webhook implemented as an AWS Lambda function and stored in a JSON file hosted in an AWS S3 bucket.
We also stored log files of all interactions with the system in a separate AWS S3 folder.

Note that to maintain data privacy, the original images were not stored. Instead, only the generated image and the pose was stored for further analysis. This was clearly communicated to the user in the consent form (which was the first screen in the application -- see Section \ref{sec:procedure}).

After a picture was taken and sent to the server for processing, the application was automatically reset once 60~seconds had passed. 
The main purpose of this timed reset was to reduce queuing for participants at the public viewing area---the bottleneck of our installation and study design---and to give more time to the subsequent interviews.
The source artworks were randomly shuffled after each image generation to provide each participant with an unbiased list of artworks to select from.
}%

\subsubsection*{Web Application 2: Image viewing}
\ok{%
The second application 
was a light-weight frontend created for viewing the resulting images.
The application used long-polling to fetch the list of generated images stored as JSON file in an AWS S3 bucket.
The research assistants could scroll back and forth through the images with keyboard arrow keys.
A red square on the bottom right of the screen alerted the assistant that newly generated images were available for viewing.
The number of new images was displayed in this red square.
The participant's pose was shown in the bottom left corner of the screen.
Generated images were viewed either on a large display in public or in private on two laptops with privacy screens, depending on which posing area was chosen by the participant.
}%


\section{Evaluation}%
\label{sec:procedure}%



\ok{%
The aim of our study was
to evaluate body prompting as a novel input modality for image generation in public settings.
We aimed to observe and understand how body prompting is used and experienced 
in the context of a well-visited public event.
Therefore, the research was conducted ``in-the-wild'', focusing on observing participants in an environment outside of the controlled laboratory, without interference 
from researchers.
The research was exploratory and participants were not given specific posing instructions.
}%


\subsection{Study Procedure}
\begin{figure}[!htb]%
\centering%
\includegraphics[width=\linewidth]{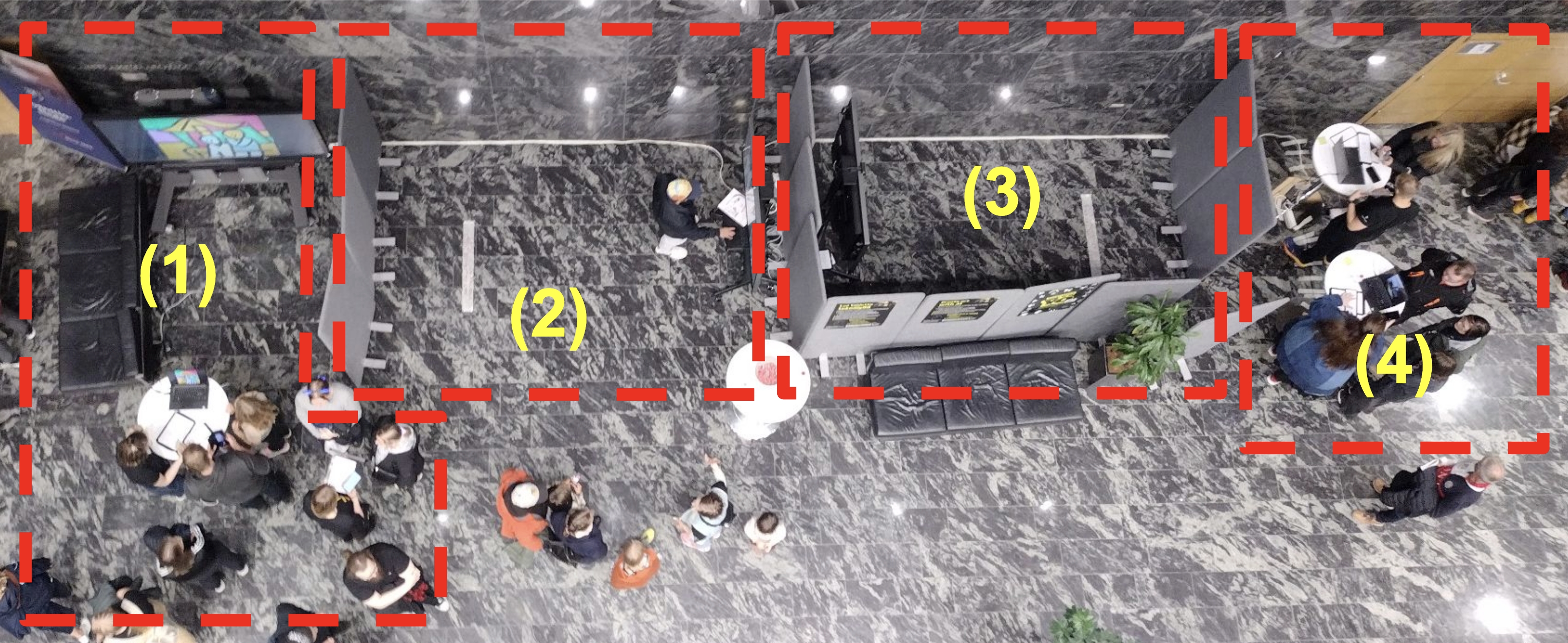}%
\caption{The study setup consisted of four separate areas: 1) the public viewing and interview area, 2) the public posing stage with touch-screen display, 3) the private photo booth with touch-screen display, and 4) the private viewing area.}%
\Description{}
\label{fig:setup}%
\end{figure}

\ok{%
The study procedure is summarized in \autoref{fig:process} and the spatial layout is depicted in \autoref{fig:setup}.
Participants were invited to walk up to the public display. The first application screen consisted of a consent form (\circnum{1}) informing the participants about the extent and use of the collected data.
Participants were specifically informed that only an anonymous, stick figure-like pose would be collected, not the original photo.
After providing consent by clicking a button, participants could scroll through the artworks, which were presented in a gallery with masonry layout (\circnum{2}).
Participants selected an artwork from the gallery by touching the screen.
Next, participants were informed that a 
timer would count down from 10~seconds. After confirming this message, 
participants could see themselves on the screen 
and assume a pose while the timer counted down to zero (\circnum{3}).
The posing was supported by showing the artwork for a brief moment before displaying the camera feed, and a small version of the artwork was additionally shown in the corner of the screen while posing.
Participants were not given any instructions on what specific pose to assume and were completely free to deviate from the source artwork.
Finally, participants were shown a short unique code from a curated list of Finnish and English words, generated with a language model, and informed to exit the booth in the final application screen (\circnum{4}).
The codes were carefully reviewed to avoid introducing bias into the study.
The short code was used to link the subsequent interview with the generated image.

\begin{figure*}[htb]%
\centering%
  \includegraphics[width=\textwidth]{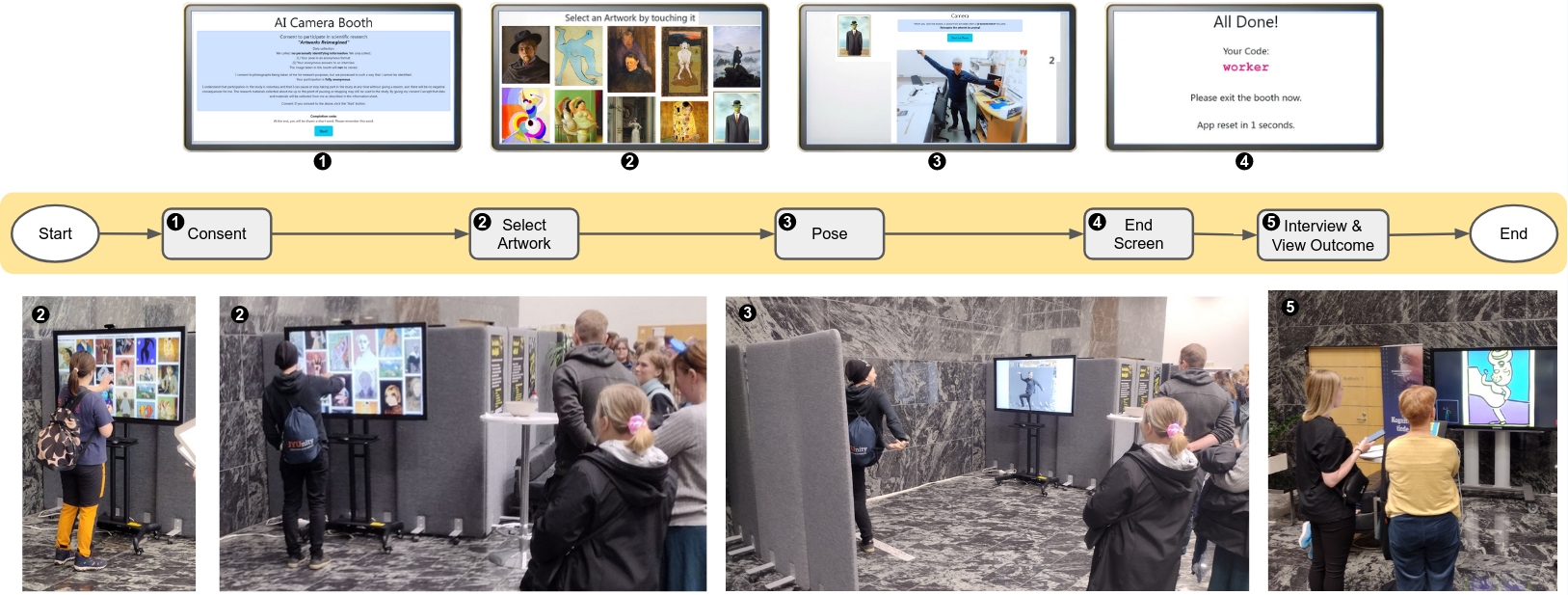}%
  \caption{The study setup of the interactive art installation, \textit{Artworks Reimagined}, guides participants through four application screens (top) in which participants give consent \circnum{1}, select an artwork \circnum{2}, \textit{body prompt} via posing in front of a camera \circnum{3}, and exit the camera booth \circnum{4}. 
  This is followed by an interview in which the results are revealed \circnum{5}.}%
  \Description{}
  \label{fig:process}%
  \label{fig:teaser}%
\end{figure*}%


Following the interaction with the installation, one of five research assistants invited the participants to a semi-structured interview
    recorded 
    using a mobile phone (\circnum{5}).
The assistants showed the participant(s) the artwork and detected body prompt during the interview and recorded participant(s)' reactions and comments in response to seeing the generated artwork.
}

\ok{%
The study was conducted in accordance with the policies of 
the University of Jyväskylä's
 Institutional Review Board (IRB) and the Finnish National Board on Research Integrity (TENK).
Children under the age of 18 were allowed to use the installation under the supervision of adults, but were excluded from the interviews. In this article, we only focus on the data collected from 79~adults who agreed to an interview following the usage of the installation.
\transition{%
The semi-structured interview questionnaire is described in the following section.%
}%
}
%
%
\subsection{Data Collection and Analysis}%
\label{sec:data-collection-analysis}%
\ok{%
We developed an interview guide consisting of 12~questions. 
Besides basic demographics, 
the research assistants asked participants about their motivation for selecting the artwork, the body prompt, and the choice of private or public posing area.
The interviewers also inquired how participants experienced body prompting and what they wanted to express with their body prompt.
Some questions were followed by open-ended 
    questions (``can you describe why?'').
A Likert-scale question captured how 
enjoyable the experience was for the participant (from 1 -- Not At All Pleasant to 5 -- Extremely Pleasant).
The interview also included the ten-item questionnaire on key personality traits, commonly referred to as ``Big Five''~\cite{BigFive}.

The generated image was revealed on the screen at the end of the interview and participants were asked to comment on how they perceived the image and whether the image expressed their original intent, thus giving insights into co-creative aspects of body prompting, and any other thoughts about their body prompt
and the generated image.
Last, participants were asked to rate and comment on the overall experience (on a Likert scale from 1 -- Not At All Enjoyable to 5 -- Extremely Enjoyable).%
}

\ok{%
The interview data was transcribed using a text-to-speech service,
translated from Finnish into English using Google Translate,
and reviewed and corrected by three research assistants
who listened to the recordings and manually corrected translation errors.
%
The 
qualitative data was analyzed following a ground\-ed theory approach \citep{Glaser_1967.pdf}.
Three authors (one Postdoc in Cognitive Science, one PostDoc in Computer Science, and one PhD student with professional experience in user experience design) discussed a preliminary coding scheme for each question as a starting point for inductive coding. The three authors then individually coded the responses of a sample of 30 participants. In a discussion following this individual coding session, the authors noticed that coding the data was straightforward 
\citep{10.1145/3359174}. That means, in all but  
extremely few cases, it was easy to identify themes in the data, and in two other questionnaire items, the responses were too diverse to identify common themes. Therefore, it was decided that one author would inductively code the interview data, paying attention to internal consistency of the data coding.
Results were then discussed in two meetings.
}%

\ok{%
Further, the generated images and body prompts (i.e., poses) were rated by three authors along the following criteria:
    a) dynamism of the source artwork (low/medium/high),
    b) dynamism of the participant's pose (low/medium/high),
    and
    c) a comparison of the source artwork and generated image in terms of whether a change in narrative took place.
For a) and b), the authors collaboratively agreed on a codebook, as listed in \autoref{tab:codebook} in Appendix~\ref{appendix:codebook}.
    With the metric of `dynamism', we intend to capture the expressiveness of the artwork and body prompt.
    A static pose was rated to be of low expressiveness, whereas a highly dynamic pose was rated as being very expressive.
For c), the authors agreed to code a change in narrative if a subject engaged in a different activity or had a different expression (e.g., static versus humorous), or if the model hallucinated artifacts that caused or strongly contributed to a change in narrative between the source image and the generated image.
After a first rating round, the ratings were discussed and each author revised their ratings for internal consistency.
}%

\ok{%
Employing Fleiss’ Kappa for inter-rater reliability assessment, our study found high agreement levels among the three raters across all three variables: source artwork dynamism ($\kappa = 0.96$), pose dynamism ($\kappa = 0.86$), and narrative change ($\kappa = 0.88$). These results underscore the raters' cohesive interpretations and a consistent and 
robust evaluation process for the studied variables.
%
A majority vote between the three authors was used to produce the final set of ratings, in cases where the authors' ratings diverged.%
}%


\ok{%
In addition, one author analyzed and coded the difference between the source artwork and the body prompt,
characterizing the latter as either being
    an imitation (re-creation) of the source artwork or an attempt to `reimagine' (reinterpret) the artwork with a body prompt that significantly differs from the source artwork.
The latter reflects the participant's intention to shift the narrative between source artwork and generated output by using a body prompt that introduces significant differences.
It was further coded whether this narrative change in the resulting image originated from the generative model (e.g., by hallucinating artifacts or facial expressions) or the user (via the body prompt), with options `model', `user', `both', and `n/a' (meaning no major change in narrative took place).%
}
%
%
%
%
%
%
\subsection{Participants}%
\label{sec:participants}%
\ok{%
Interviews were conducted with 79~visitors (henceforth called participants) after they used the installation.
Participants were on average 37~years old ($Median=35$ years, $SD=13.5$ years, $Max=68$ years) and
included 46~women, 32~men, and one non-binary participant.


Participants' Big Five personality traits---Openness, Conscientiousness, Extraversion, Agreeableness, and Neuroticism---were assessed on five-point Likert scales.
Participants' above average openness scores ($Mean=3.56$, $SD=0.83$) suggest a general tendency towards creativity and openness to experience within the cohort.
Conscientiousness followed closely ($Mean = 3.51$, $SD=0.72$), indicating a high level of reliability and organization among participants.
Extraversion and Agreeableness scores were slightly lower ($Means = 3.20$ and $3.19$, respectively, $SDs = 0.56$ and $0.72$), pointing to moderate levels of sociability and empathy.
Neuroticism received the lowest average score ($Mean = 2.93$, $SD=0.76$), suggesting a lower degree of emotional instability and stress sensitivity in the group. These findings offer a nuanced view of the personality composition within the participant sample, highlighting a 
tendency towards openness and conscientiousness, and providing context for our exploration into the behaviors of the study cohort.%

Due to the study being conducted in the field without control over the number of participants per condition (i.e., group versus solo participation, body prompting 
strategy, artwork choice, and public versus private posing area), we focus on describing the images in Section~\ref{sec:artworks} and qualitatively analyzing the interview data in sections \ref{sec:user-experience}--\ref{sec:observations}.
For a rich description, we contextualize participant quotes with the codes listed in
\autoref{tab:participantcodes}.

\begin{table}[!htb]%
\caption{Contextual information on participants.
}%
\Description{}
\label{tab:participantcodes}%
\small%
\centering%
\begin{tabular}{lcccc}%
\toprule
   Code & Participants & Posing Area & Posing Strategy & Participation \\
\midrule
    A & 22 & Public & Reimagine & Group \\
    B & 13 & Private & Reimagine & Group  \\
    C & 12 & Public & Reimagine & Alone \\
    D & 9 & Private & Reimagine & Alone  \\
    E & 8 & Public & Imitate & Alone  \\
    F & 6 & Private & Imitate & Alone \\
    G & 5 & Private & Imitate & Group \\
    H & 4 & Public & Imitate & Group \\
\bottomrule%
\end{tabular}%
\end{table}%

For participating groups, interviews were held with one group member, but other group members were allowed to comment.
For simplicity, we refer to both groups and single participants as `participant' in this work.
}%

\subsection{Results}%
\label{sec:results}%
\ok{%
In this section, we describe the images generated at the event (Section \ref{sec:artworks}) and how participants experience the installation and co-creation with the AI (Section \ref{sec:user-experience}).
We further describe the key choices and behavior of participants and different body prompting strategies (Section~\ref{sec:behavior}).
Last, we report on the differences between the public and private setting (Section~\ref{sec:private})
and miscellaneous observations (Section \ref{sec:observations}).
This section mainly reports on general trends, while specific outstanding instances are highlighted in Section~\ref{sec:specificinstances}.
}%

\subsubsection{RQ1: Generated images}
\label{sec:artworks}
\ok{%
In total, 172 images were created during the six-hour event (112~images in public and 60~images in the private booth).
Due to the diversity of source artworks, the body prompts are difficult to compare.
In this section, we provide a brief descriptive overview of the 172~generated images and the body prompts. 
A gallery of all generated images is available at
\href{https://www.artworksreimagined.com/}{https://artworksreimagined.com}.%
}%

\newlength{\imgw}%
\setlength{\imgw}{.105\textwidth}%
\begin{figure*}[!htb]%
  \centering%
  \includegraphics[width=\imgw]{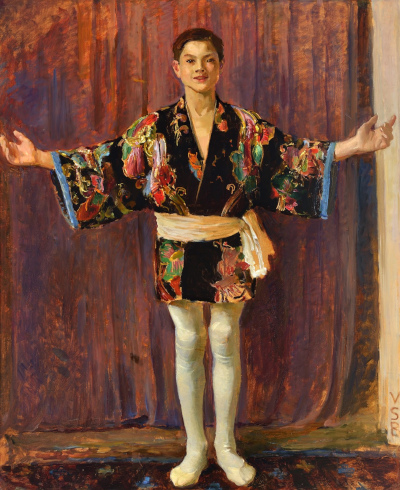}
  \includegraphics[width=\imgw]{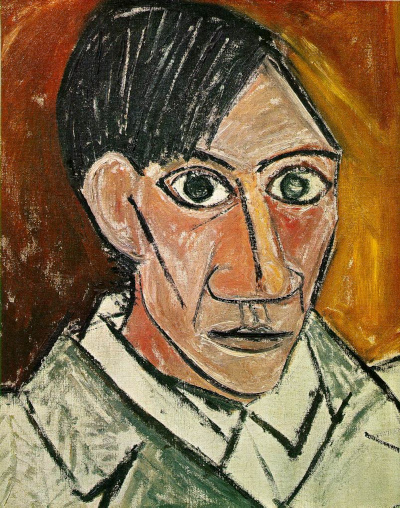}
  \includegraphics[width=\imgw]{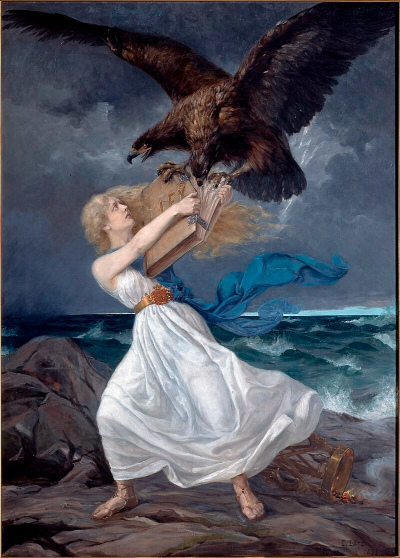}
  \includegraphics[width=\imgw]{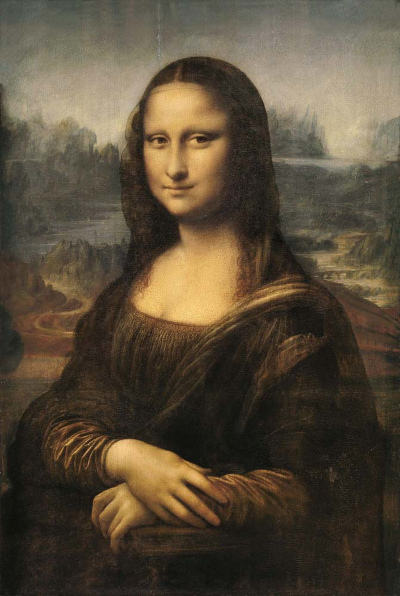}
  \includegraphics[width=\imgw]{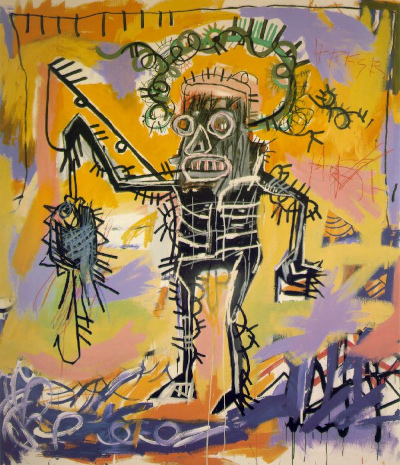}
  \includegraphics[width=\imgw]{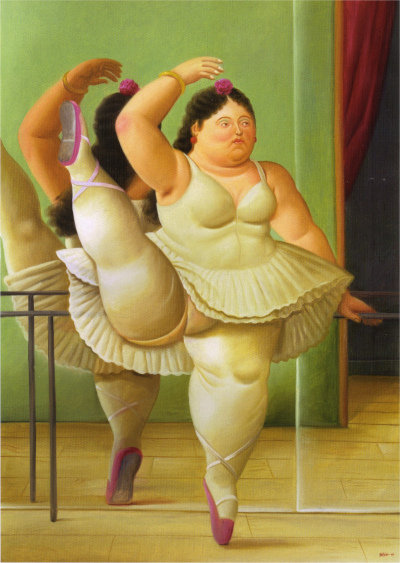}
  \includegraphics[width=\imgw]{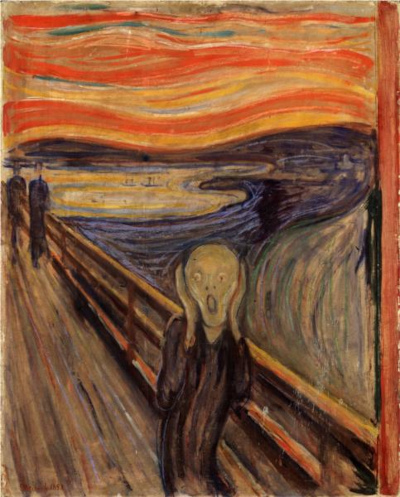}
  \includegraphics[width=\imgw]{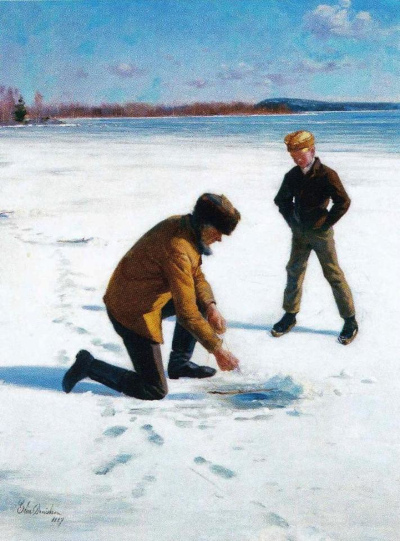}%
  \includegraphics[width=\imgw]{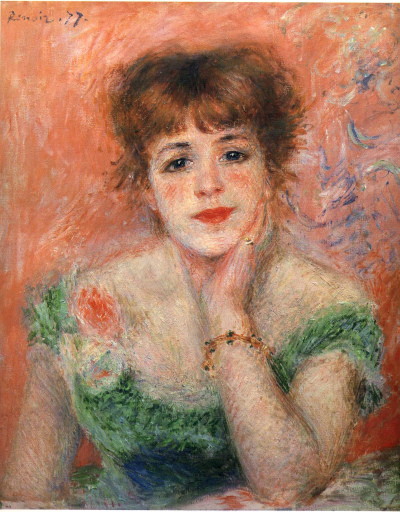}%
  \\
  \includegraphics[width=\imgw,height=\imgw]{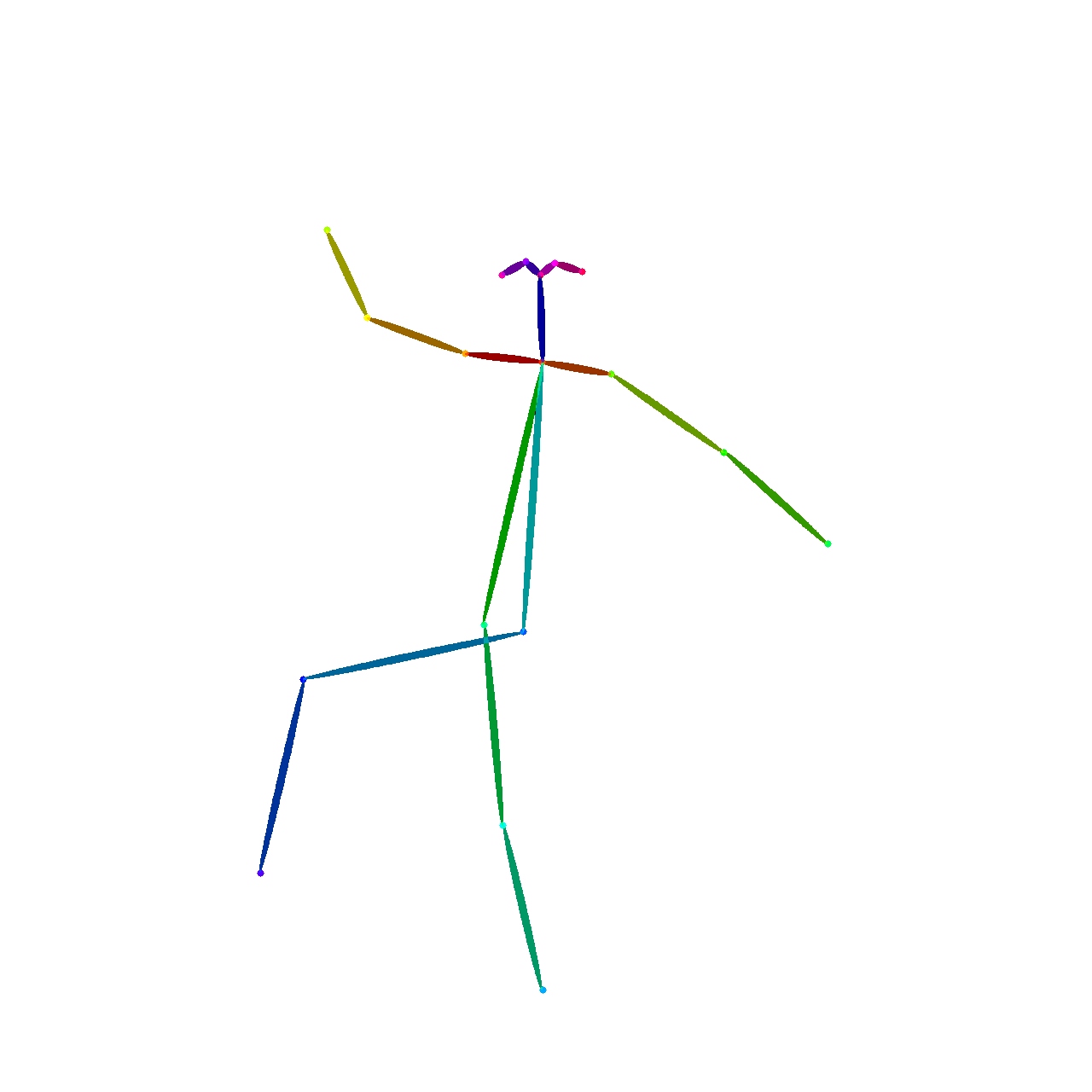}
  \includegraphics[width=\imgw,height=\imgw]{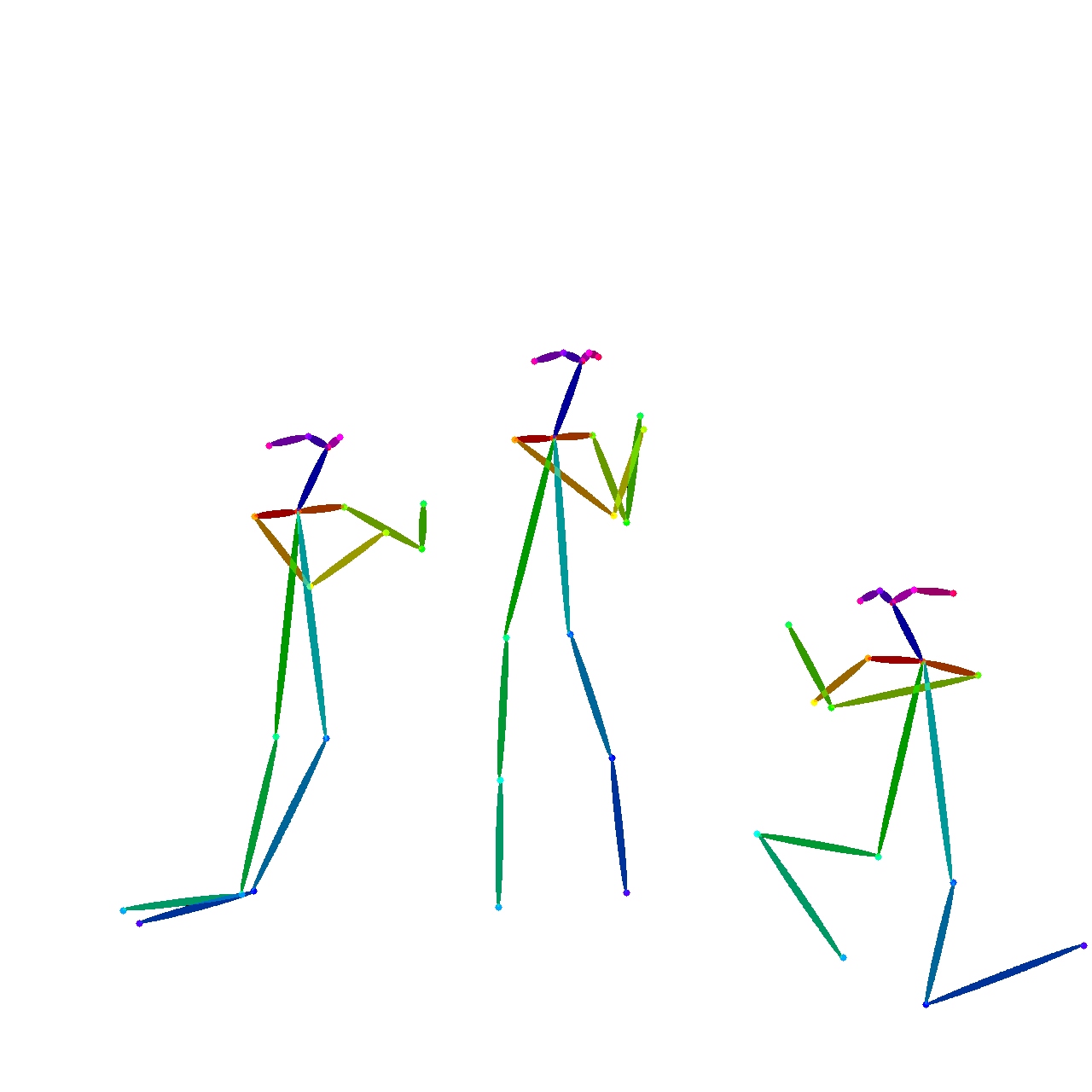}
  \includegraphics[width=\imgw,height=\imgw]{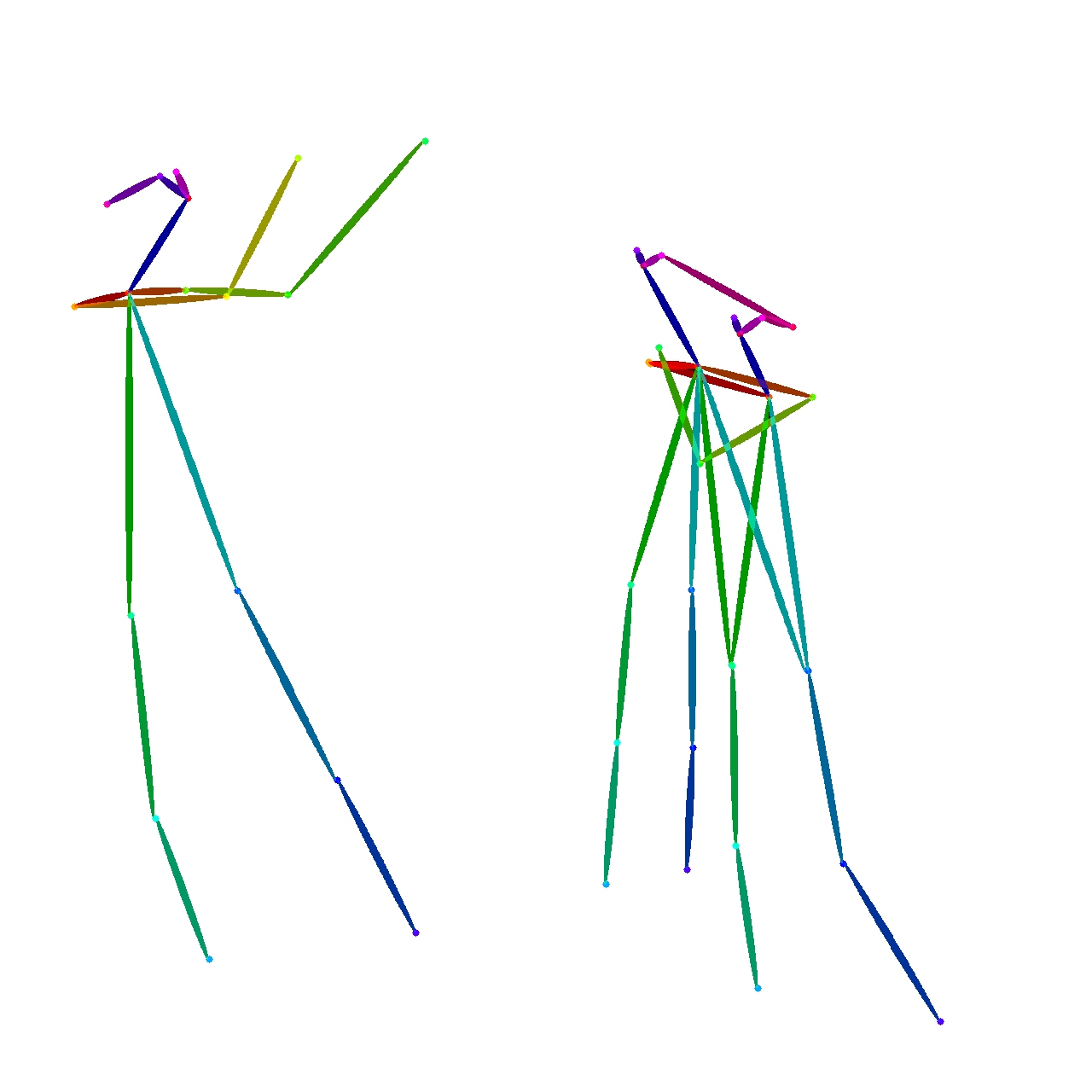}
  \includegraphics[width=\imgw,height=\imgw]{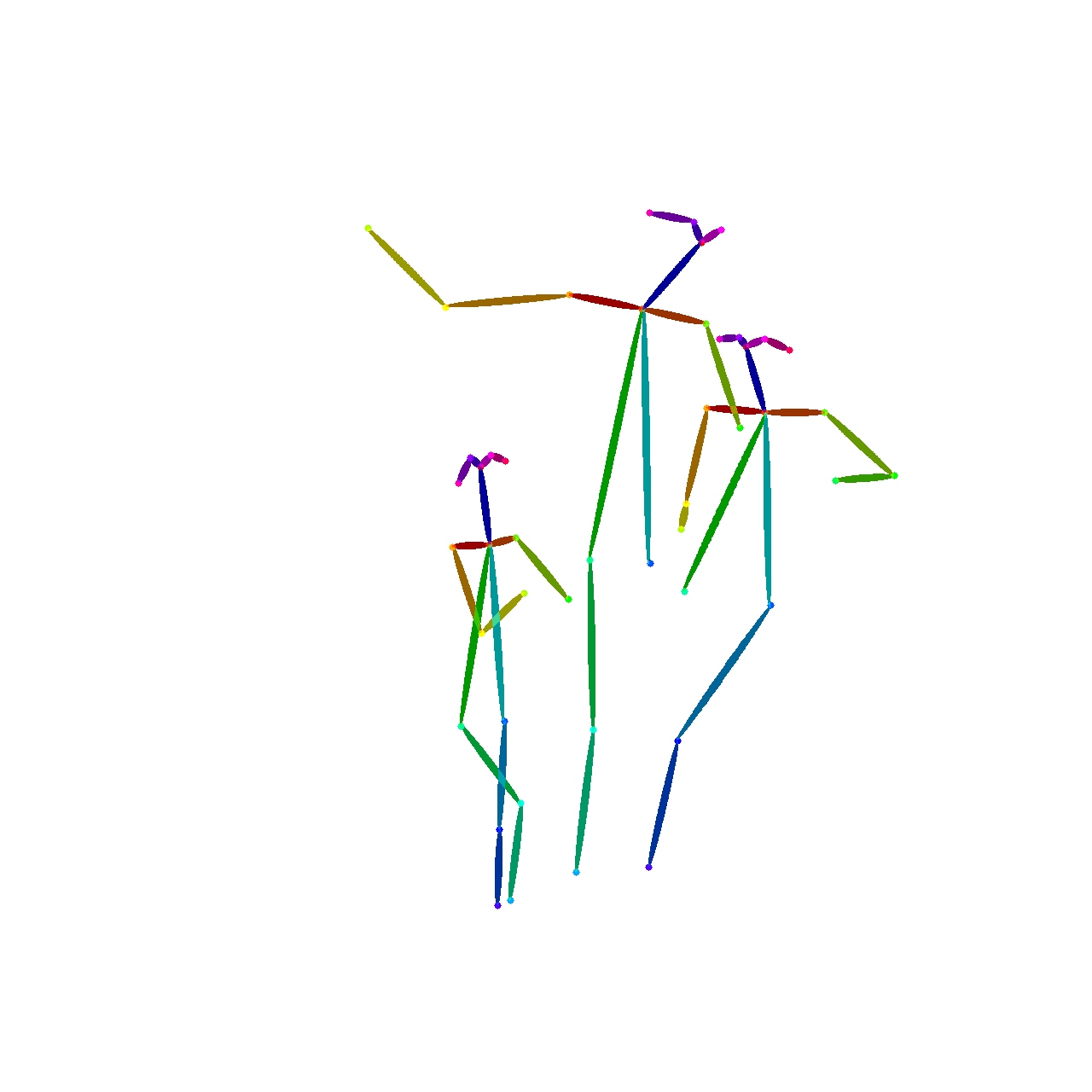}
  \includegraphics[width=\imgw,height=\imgw]{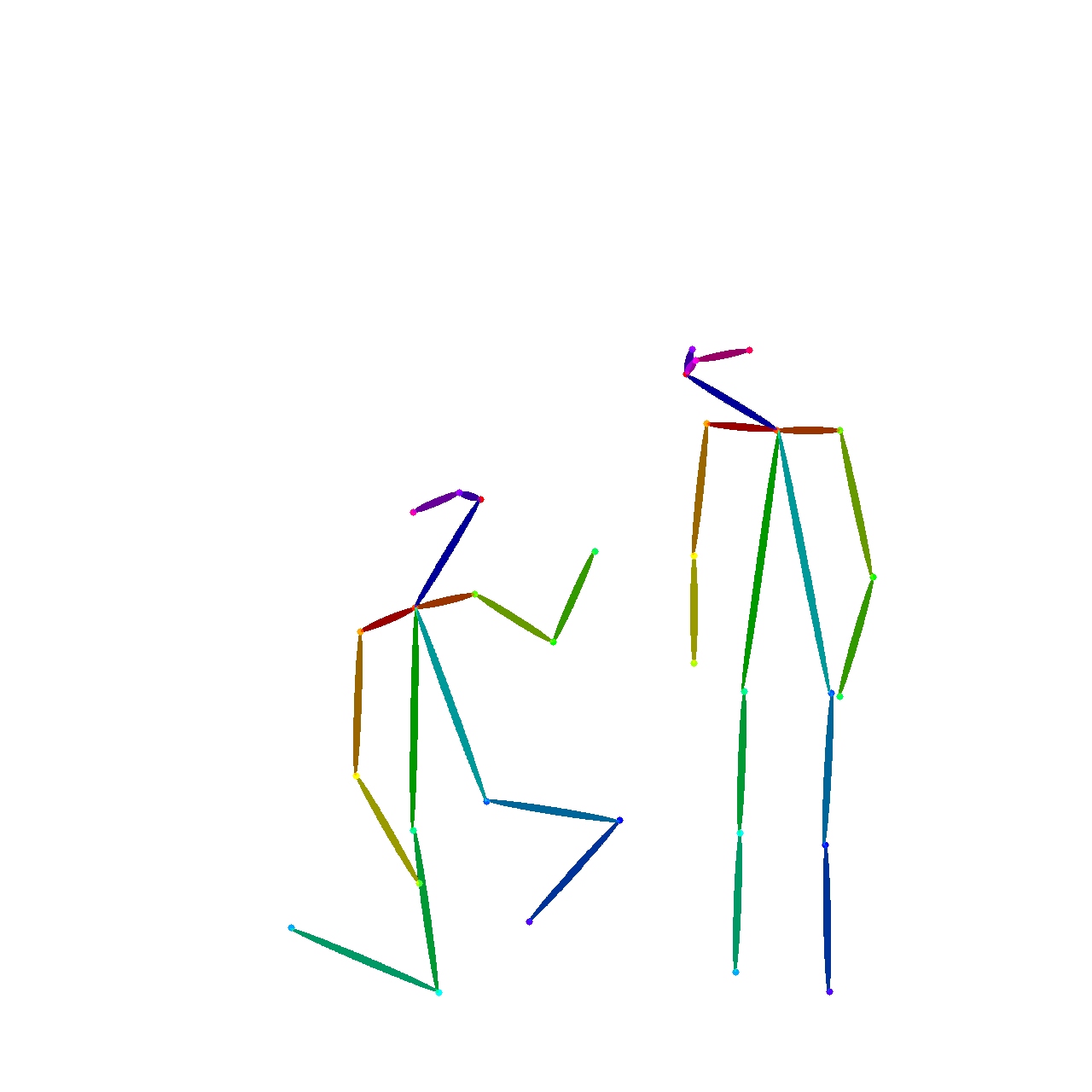}
  \includegraphics[width=\imgw,height=\imgw]{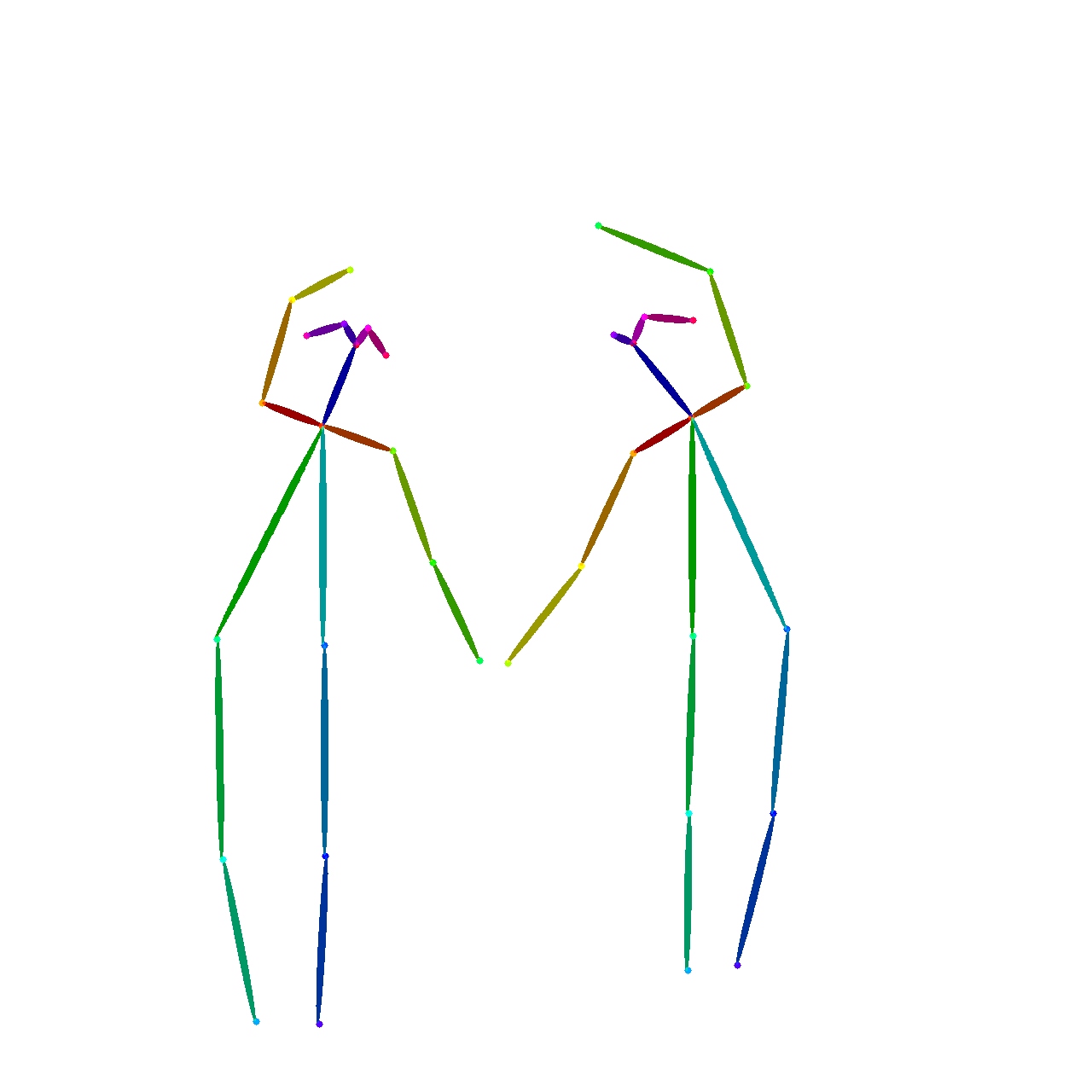}
  \includegraphics[width=\imgw,height=\imgw]{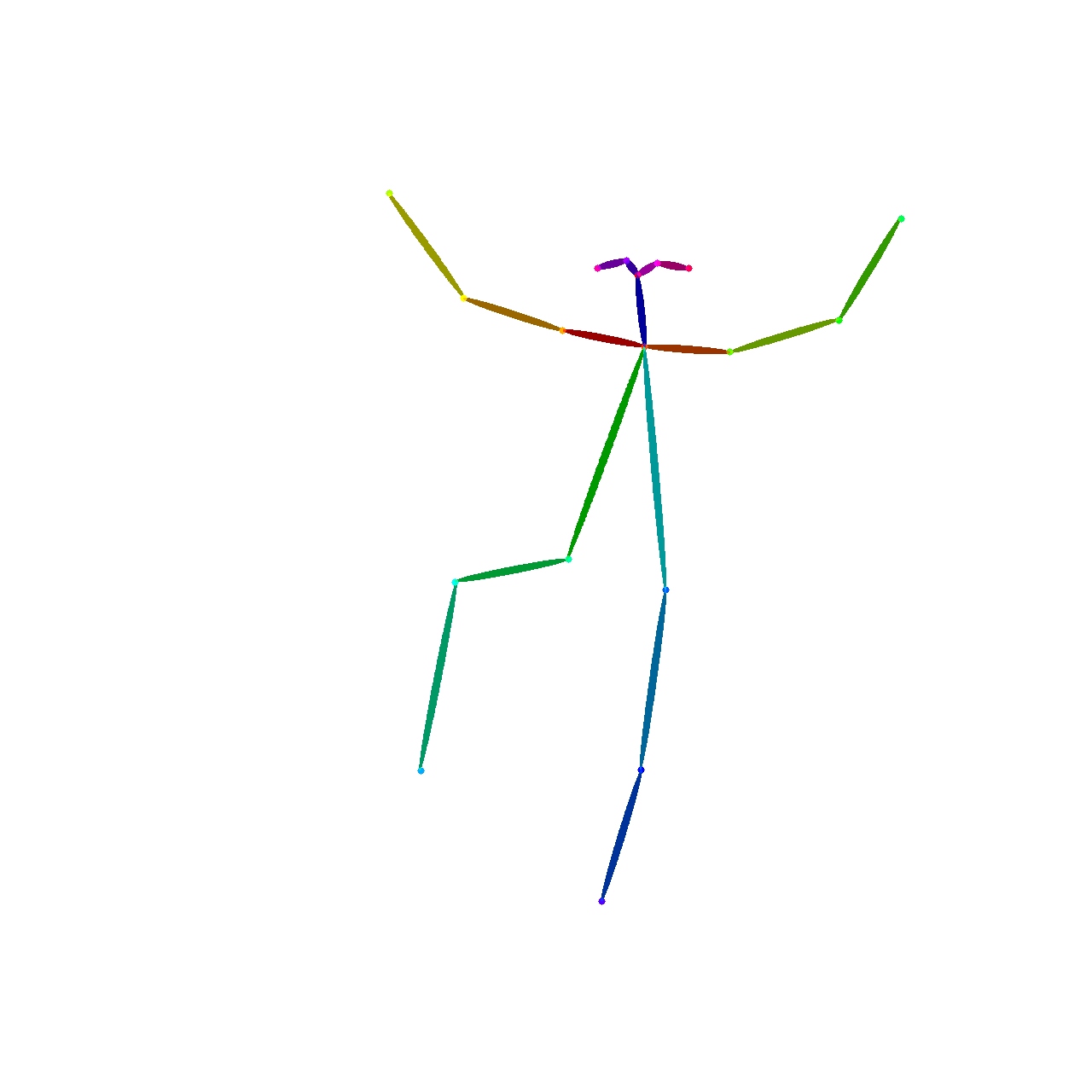}
  \includegraphics[width=\imgw,height=\imgw]{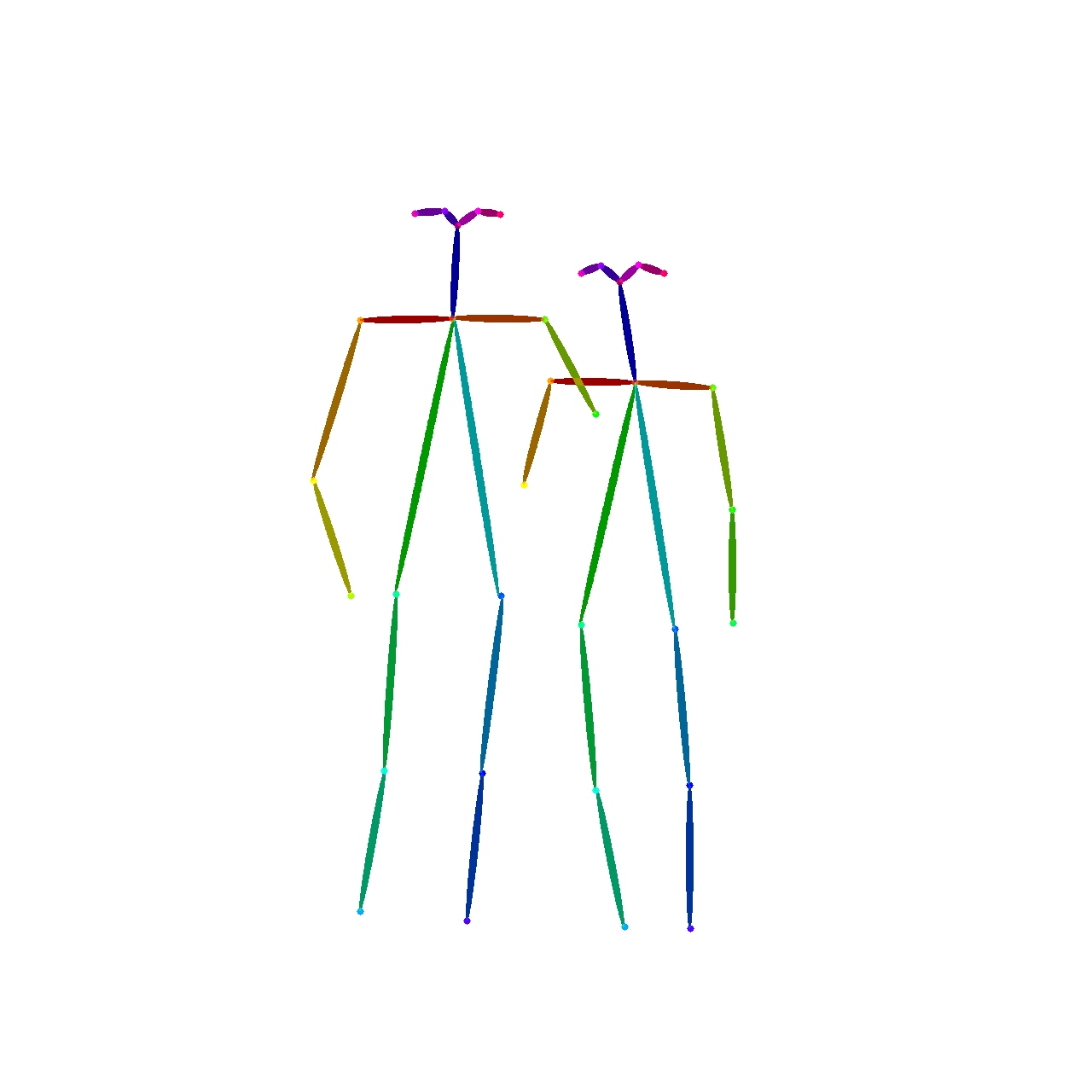}%
  \includegraphics[width=\imgw,height=\imgw]{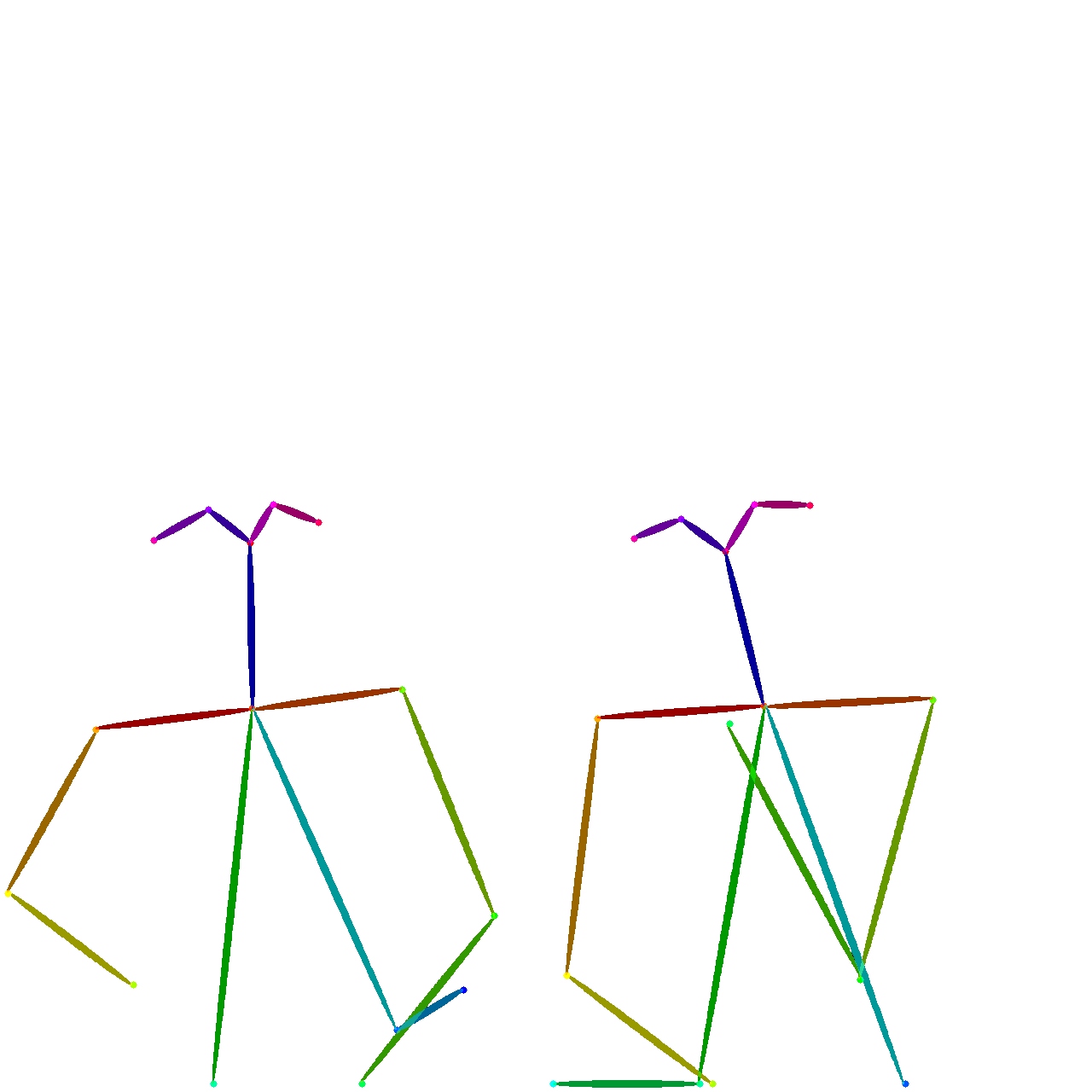}%
  \\[1mm]
  \includegraphics[width=\imgw,height=\imgw]{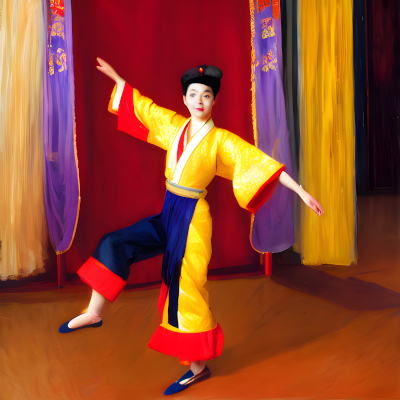}
  \includegraphics[width=\imgw,height=\imgw]{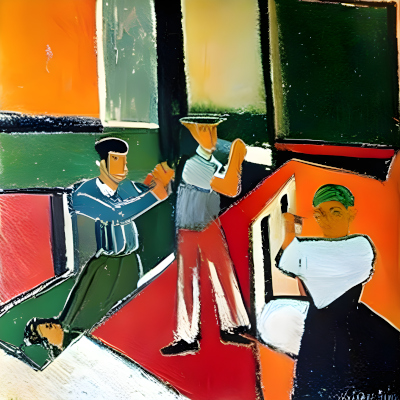}
  \includegraphics[width=\imgw,height=\imgw]{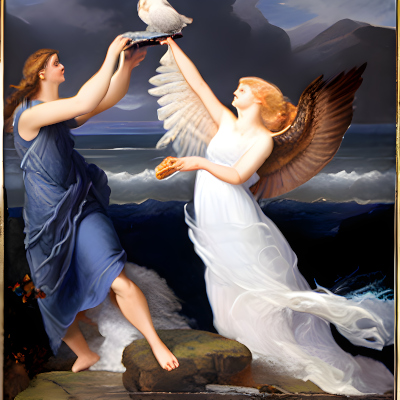}
  \includegraphics[width=\imgw,height=\imgw]{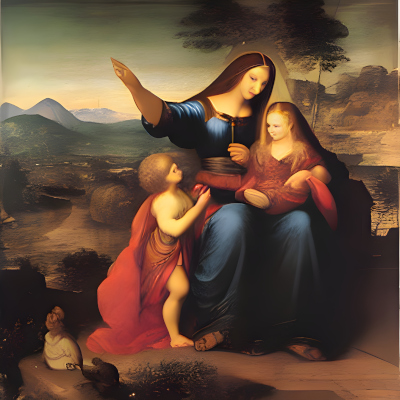}
  \includegraphics[width=\imgw,height=\imgw]{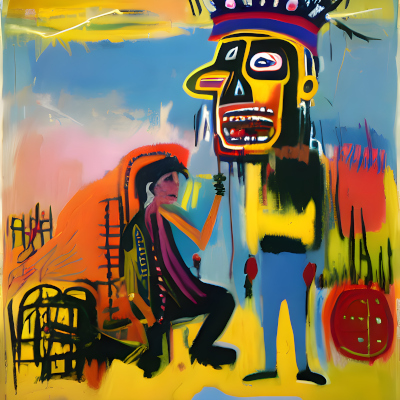}
  \includegraphics[width=\imgw,height=\imgw]{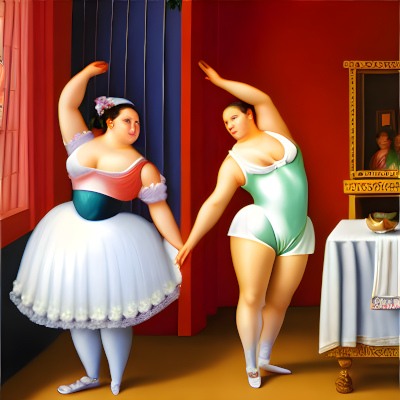}
  \includegraphics[width=\imgw,height=\imgw]{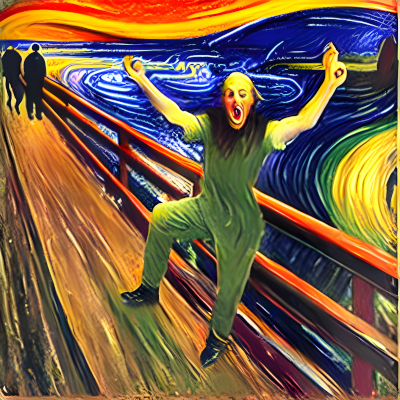}
  \includegraphics[width=\imgw,height=\imgw]{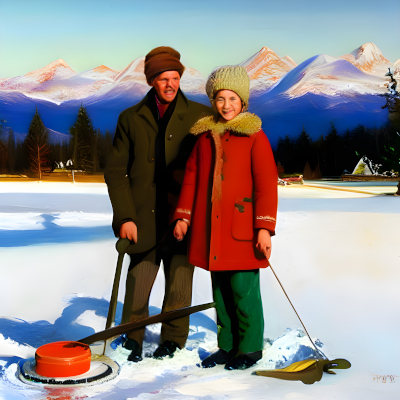}
  \includegraphics[width=\imgw,height=\imgw]{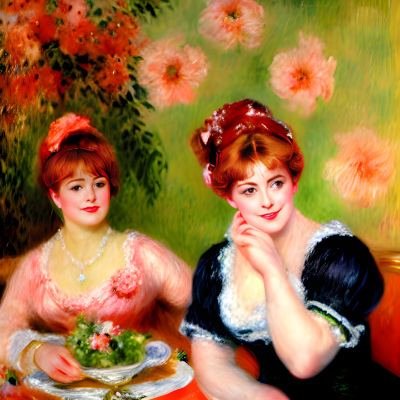}
  %
  %
  %
  %
  %
  %
  %
  \caption{Example of body prompts (middle row) and resulting images (bottom row) created during the event, together with the respective source artworks (top row). An online gallery of all images is available at
  \iftoggle{anonymous}{%
https://\textbf{[ANON-LINK]}%
}{%
\href{https://www.artworksreimagined.com/}{https://artworksreimagined.com}%
}.}%
  \Description{Example of body prompts and resulting generated images.}%
  \label{fig:artworks}%
\end{figure*}%


\ok{%
Overall, the style of the source artwork was reproduced well, to a degree where it was clearly recognizable which source artwork was used as style reference (see \autoref{fig:artworks}).
However, the generated images captured the source artwork to varying degrees, leaving room for interpretation and surprise.

The body prompts had varying degrees of expressiveness.
Some participants simply stood motionless in front of the camera in a neutral pose with arms dangling beside their hips.
Other participants were more dynamic and expressive, raising limbs into the air.
A few participants went to great lengths to produce their body prompts (as depicted in \autoref{fig:examples}, bottom center).
The system's 
    output accurately reproduced the number of posing subjects.
Overall, there were no noticeable differences between the body prompts performed in public and private, with a few exception (described in Section \ref{sec:specificinstances}).
}%

\ok{%
We classified whether a change in narrative had taken place between source artwork and generated image.
In almost half of the generated images (46.8\%) a change 
originated from the participant's body prompt, which means the 
pose introduced significant 
modifications to the source image, shifting the narrative.
In another 29.1\% of the generated images, no shift in narrative took place and the new generated image appeared to be a reproduction of the source artwork's original idea, 
    with only some minor changes.
}%

\ok{%
About a quarter of the participants (24.1\%) encountered surprising generative artifacts.
Co-creation with AI can involve surprising hallucinations by the generative model.
  With hallucinations, we specifically refer to artifacts introduced by the generative model that cannot be found in the source artwork or in the participant's body prompt, often resulting in a shift in narrative between the source artwork and the generated image.
  In our study, hallucinatory artifacts included door frames, wooden sticks, fields with houses in the distances, animals, and more.
  These artifacts appeared in the generated image without being specifically prompted or being part of the source artwork.
  However, these generated elements in the image did not necessarily deviate from the narrative portrayed in the source artwork.
  Instead, they sometimes complemented what was already in the source artwork, or they complemented the body prompt.
}%

\ok{%
About 8\% of the participants encountered surprising artifacts originating from the generative model that clearly shifted the narrative. Another 16.2\% of participants encountered images where the shift in narrative originated both from the model and the participant. These were often images where the participant's pose sbstantially deviated from the source artwork, with the AI adding surprising elements, such as additional characters or changes to the subjects.
In other instances, the model added minor hallucinatory elements that still resulted in a change of narrative.
    For instance, in the generated image in \autoref{fig:examples}, bottom right, the model appears to have added COVID face masks to the subjects' faces. These masks were not present in the source artwork.
    In another funny example, the generative model seemingly changed the gender of a subject from a blonde woman to bald man (see \autoref{fig:examples}, middle row, center).
}%



\newcommand{\imgwrapperNew}[3]{%
\begin{tikzpicture}
    \node[anchor=south west,inner sep=0] (largeImage) at (0,0) {\includegraphics[width=.20\textwidth]{#1}};
    \begin{scope}[x={(largeImage.south east)},y={(largeImage.north west)}]
        \node[anchor=north west,inner sep=0] at (1,1) {\includegraphics[width=.09\textwidth]{#2}};
        \node[anchor=south west,inner sep=0] at (1,0) {\includegraphics[width=.09\textwidth]{#3}};
    \end{scope}
\end{tikzpicture}%
}

\begin{figure*}[!htb]%
  \centering%
  \imgwrapperNew{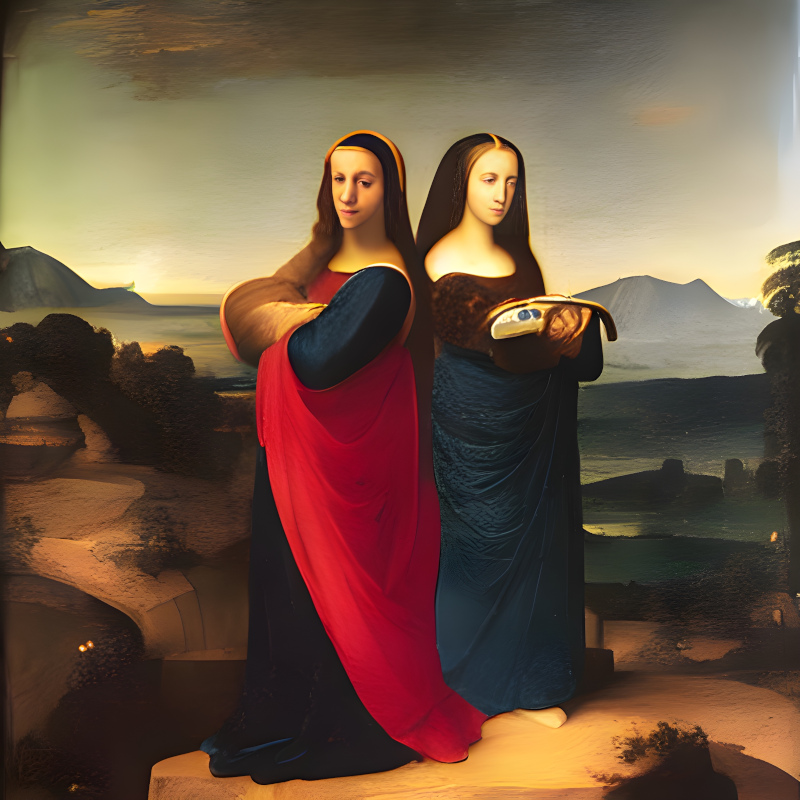}{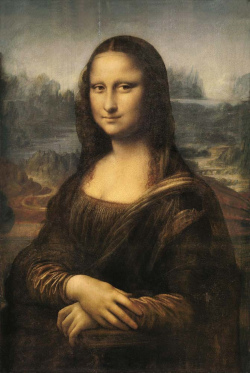}{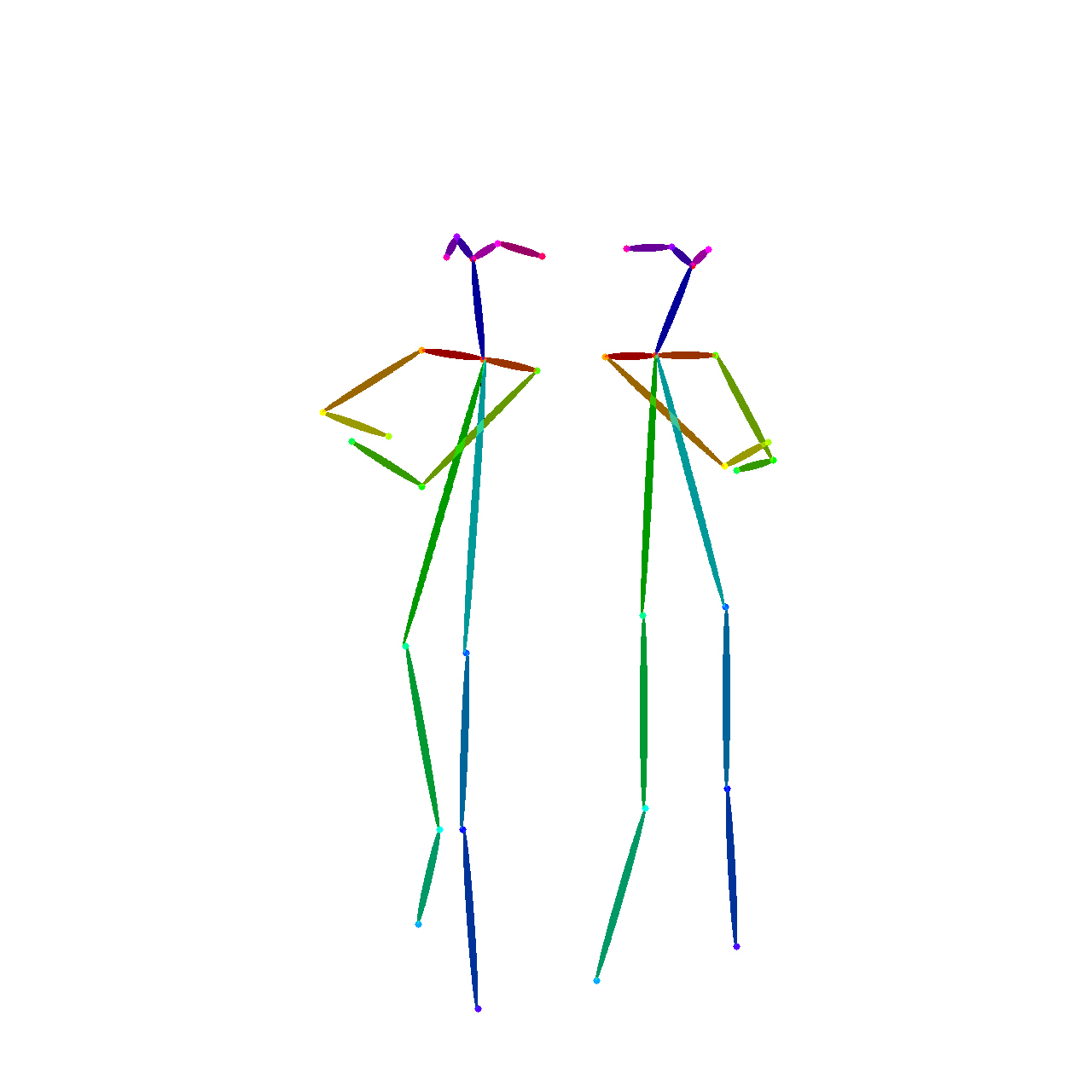}
  \hspace{1mm}
  \imgwrapperNew{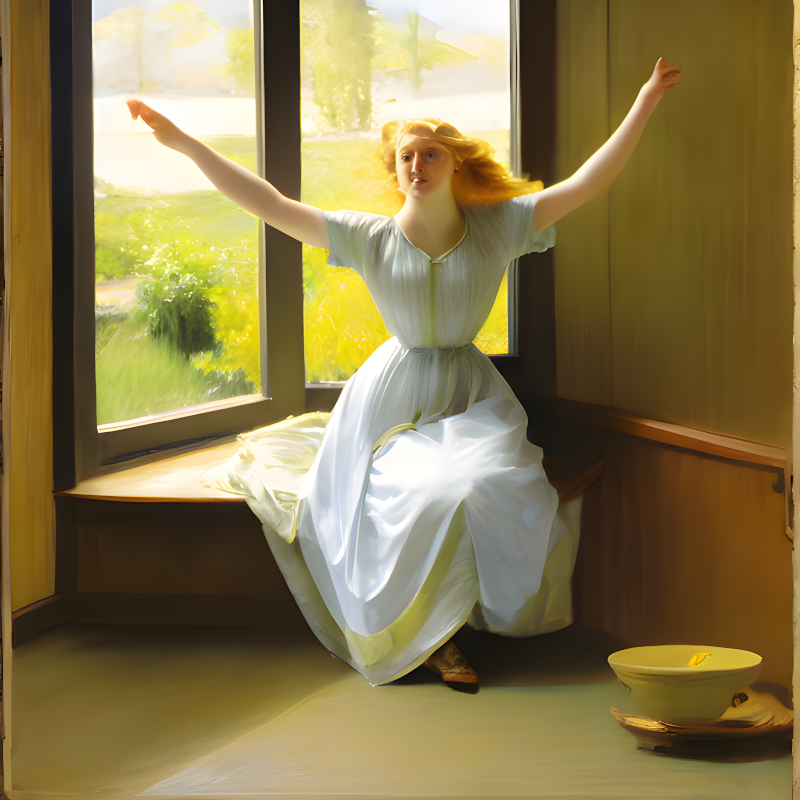}{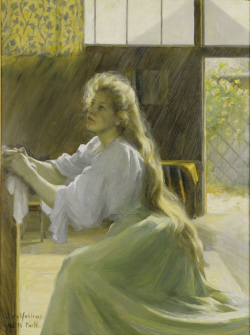}{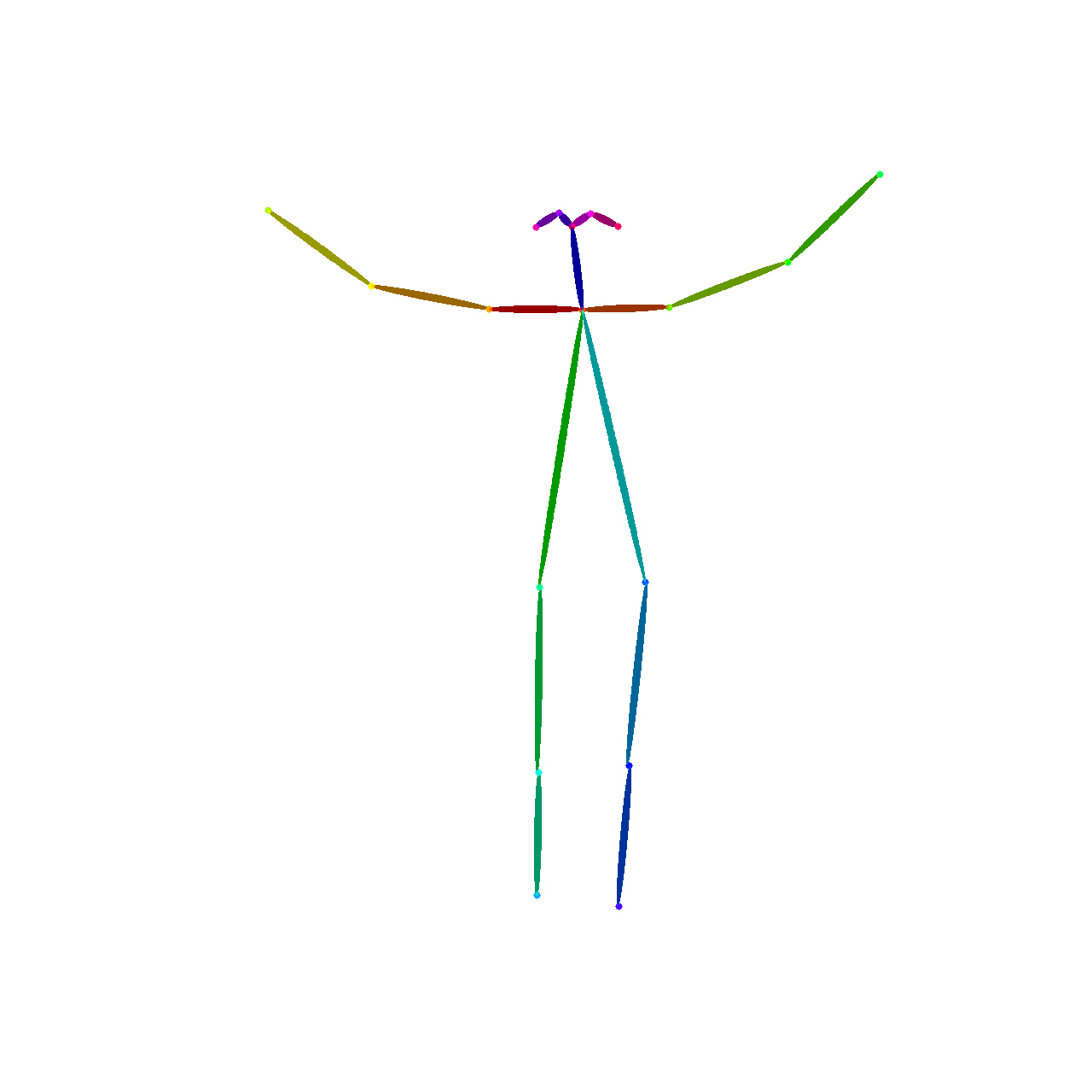}
  \hspace{1mm}
  \imgwrapperNew{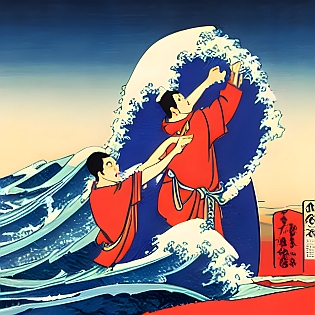}{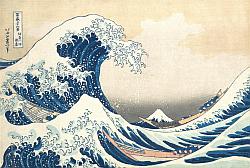}{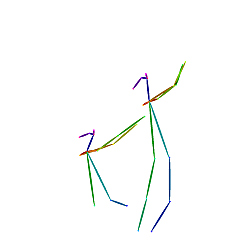}
  \\[2mm]
  \imgwrapperNew{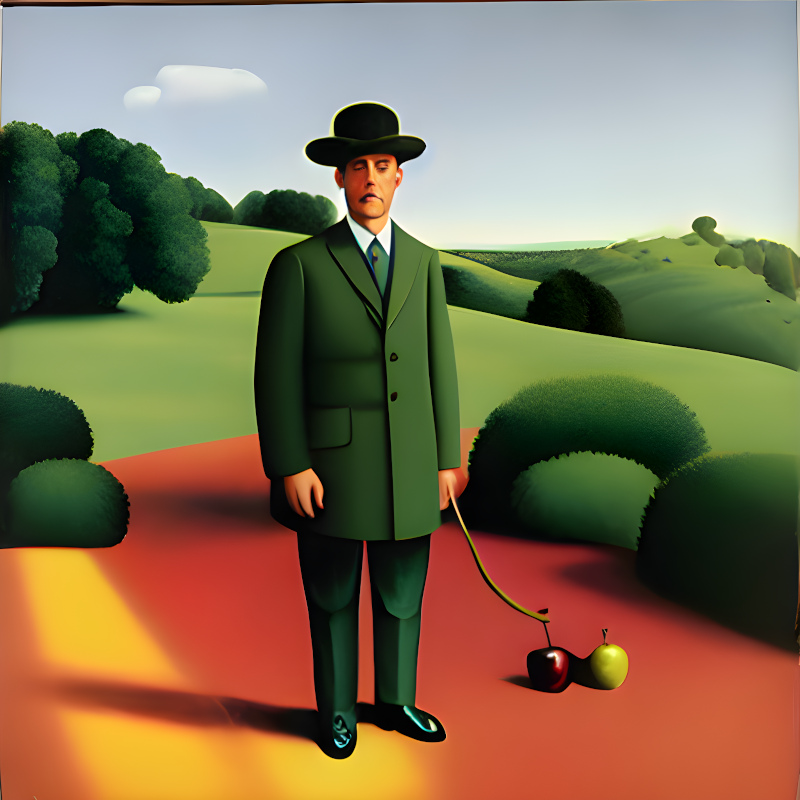}{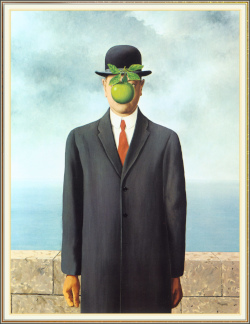}{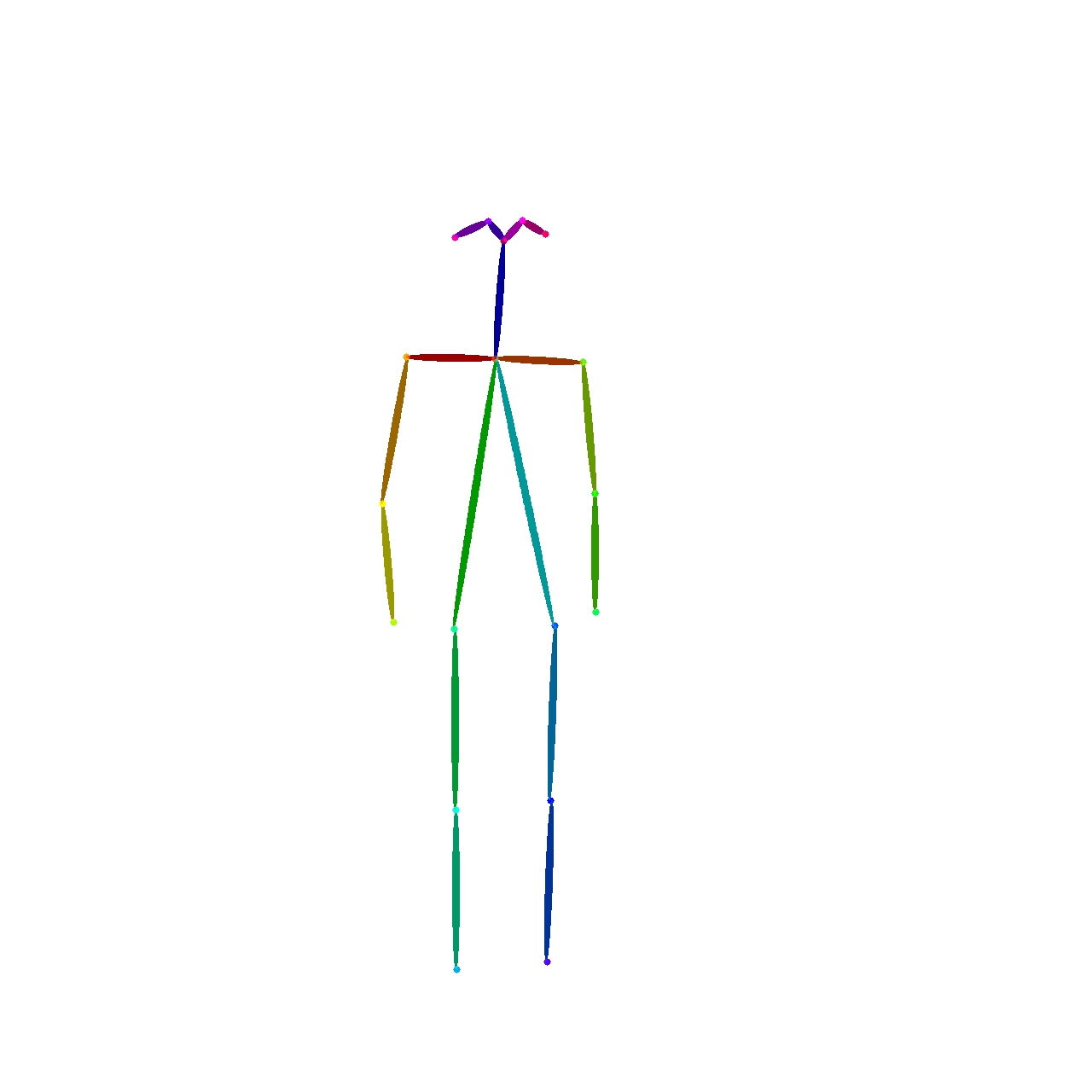}
  \hspace{1mm}
  \imgwrapperNew{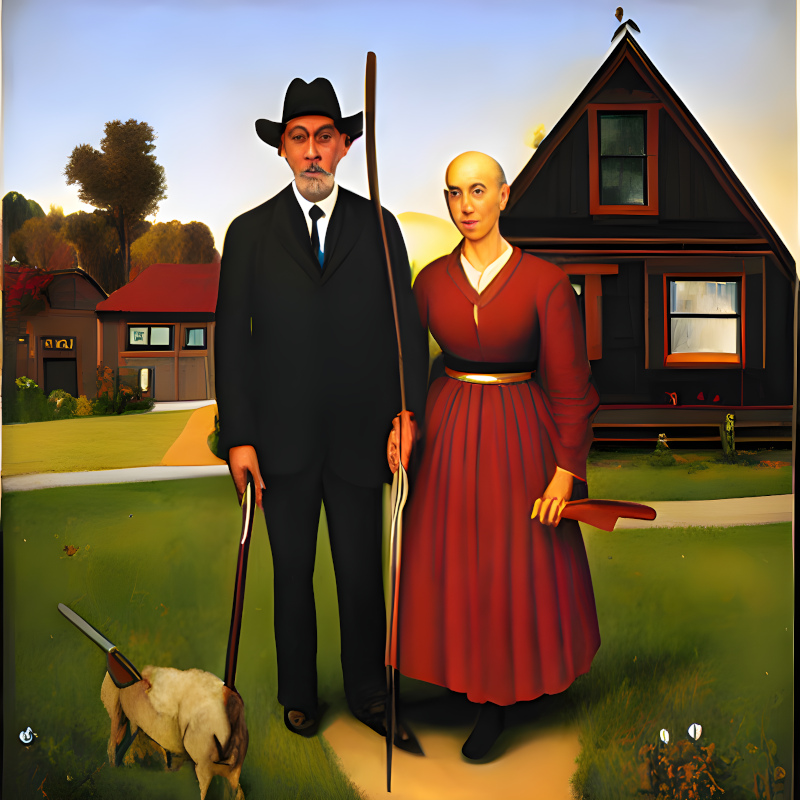}{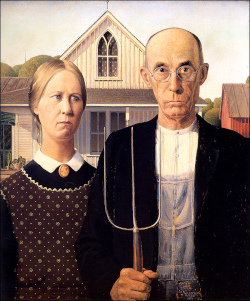}{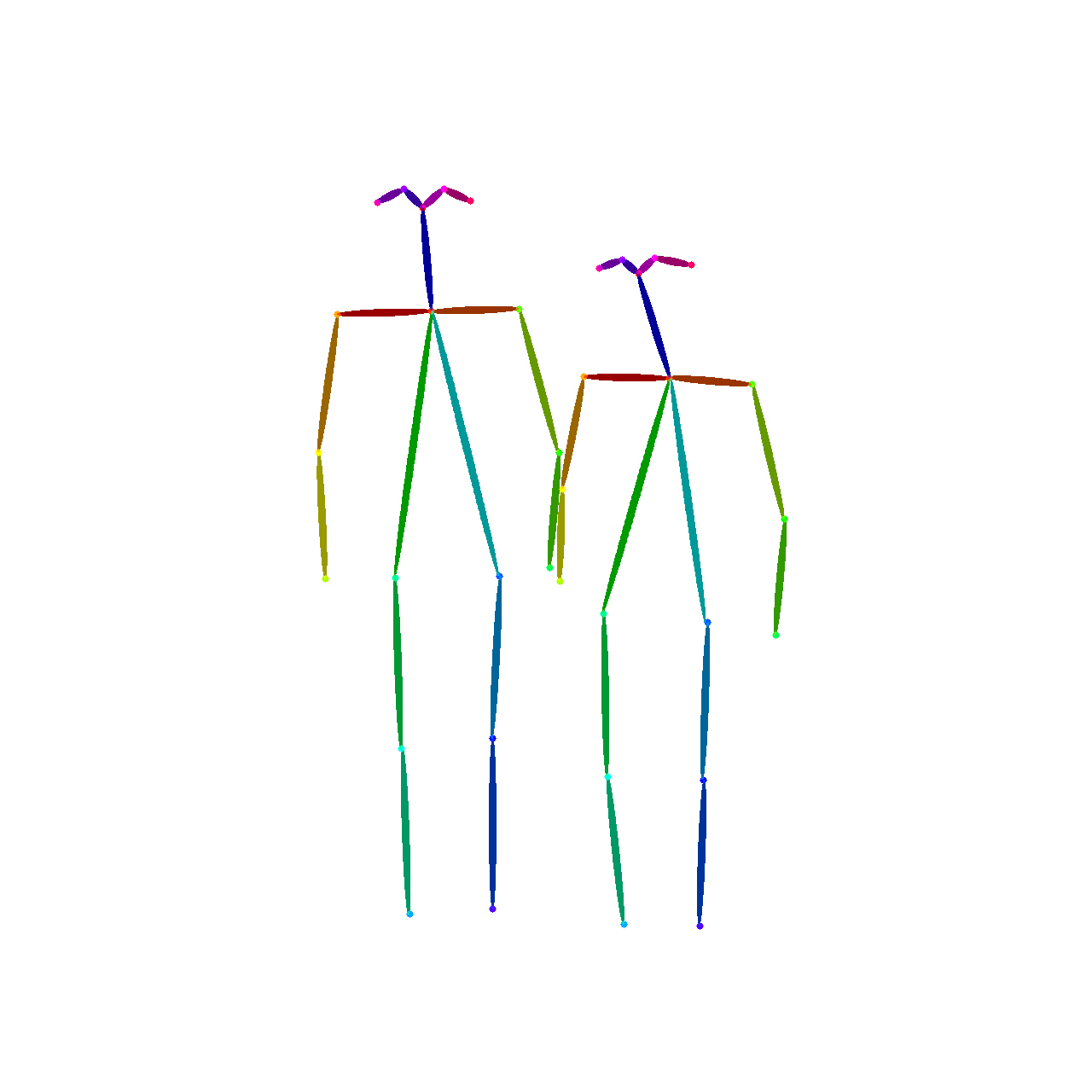}
  \hspace{1mm}
  \imgwrapperNew{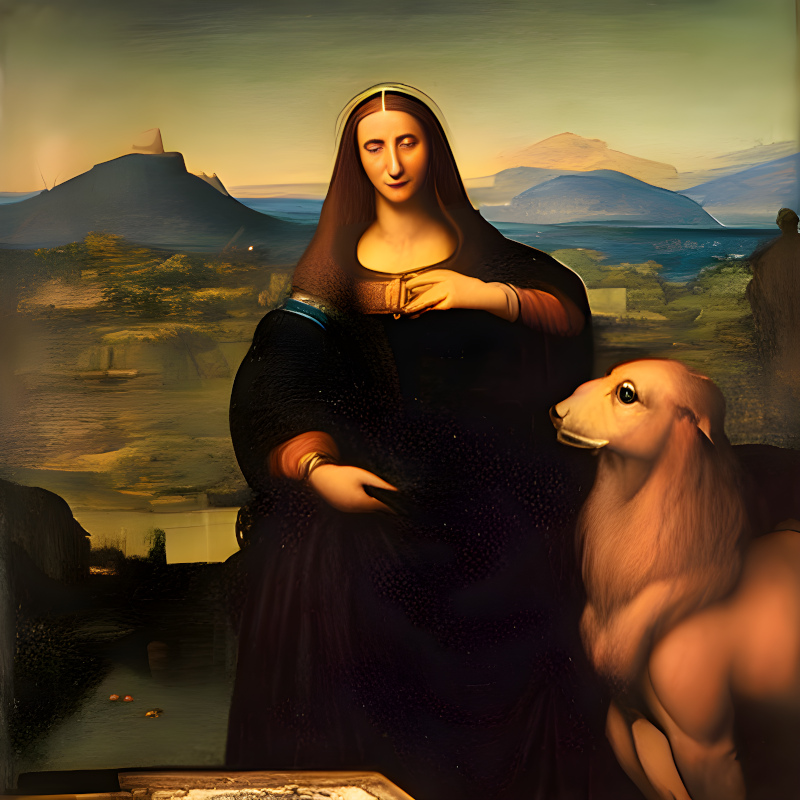}{figures-MICRO/artworks/inputs/wikiart/mona-lisa-c-1503-1519.jpg}{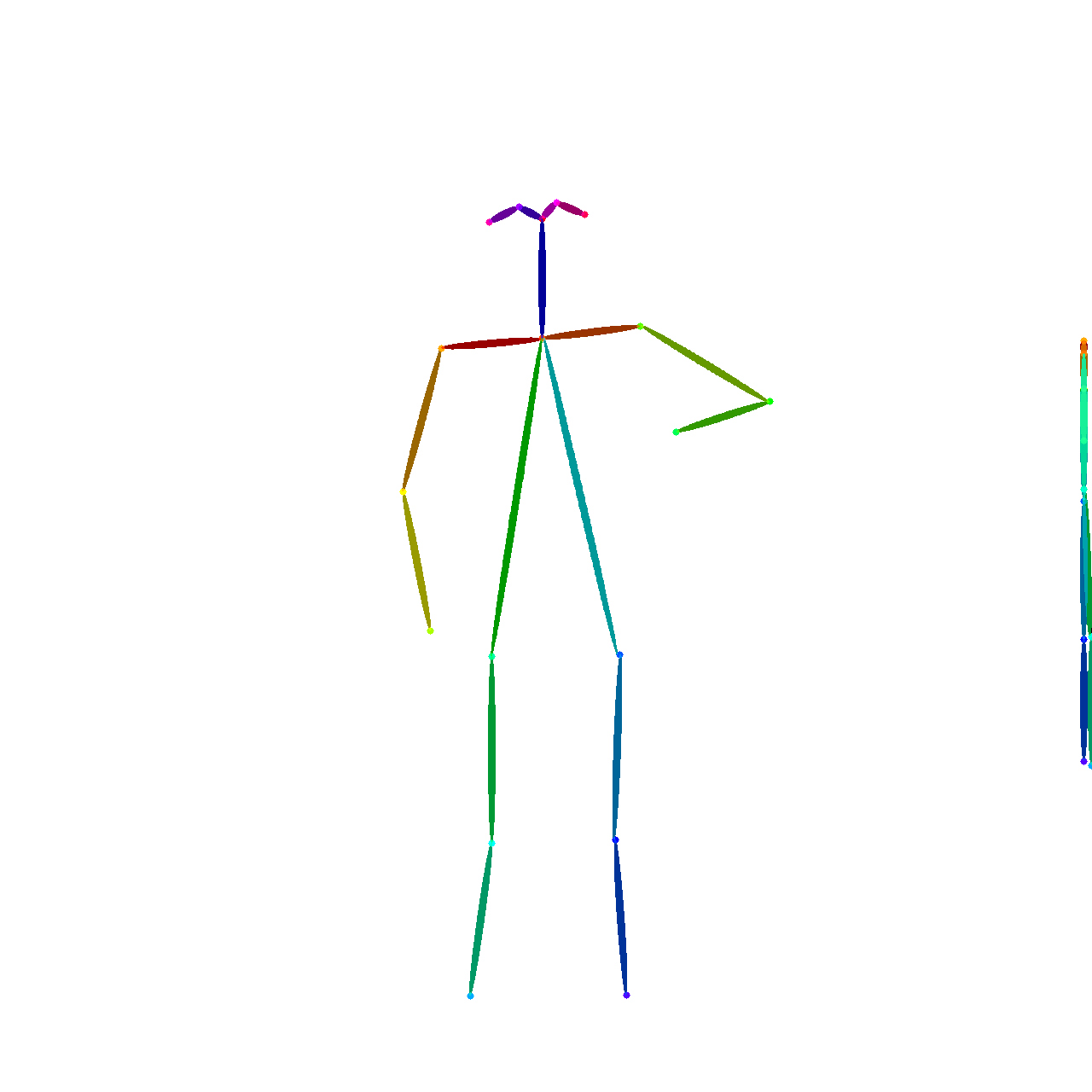}
  \\[2mm]
%
    \imgwrapperNew{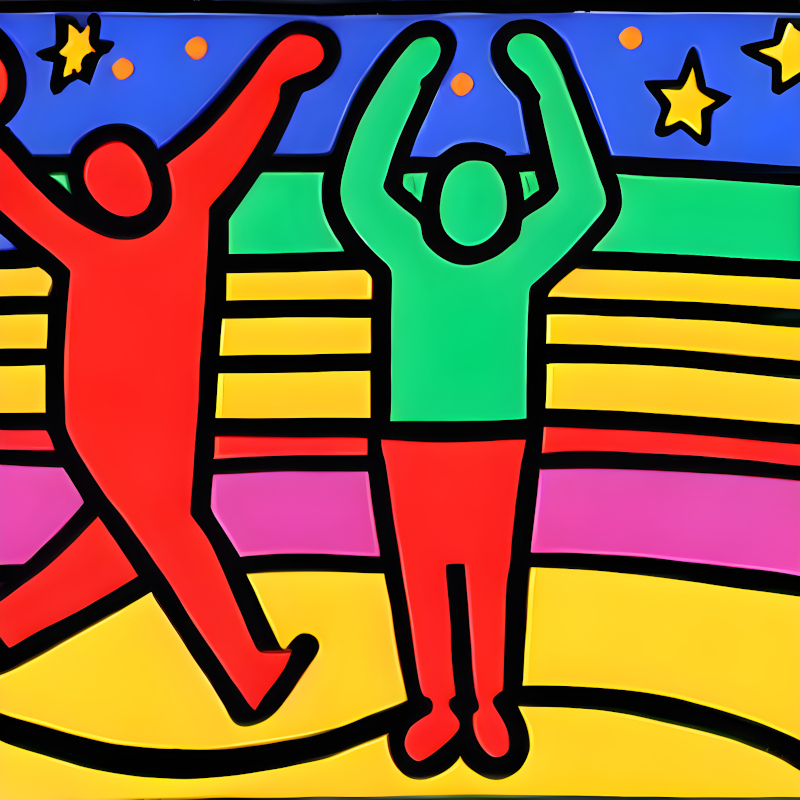}{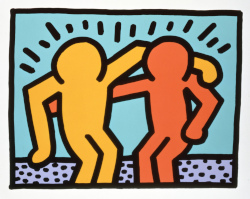}{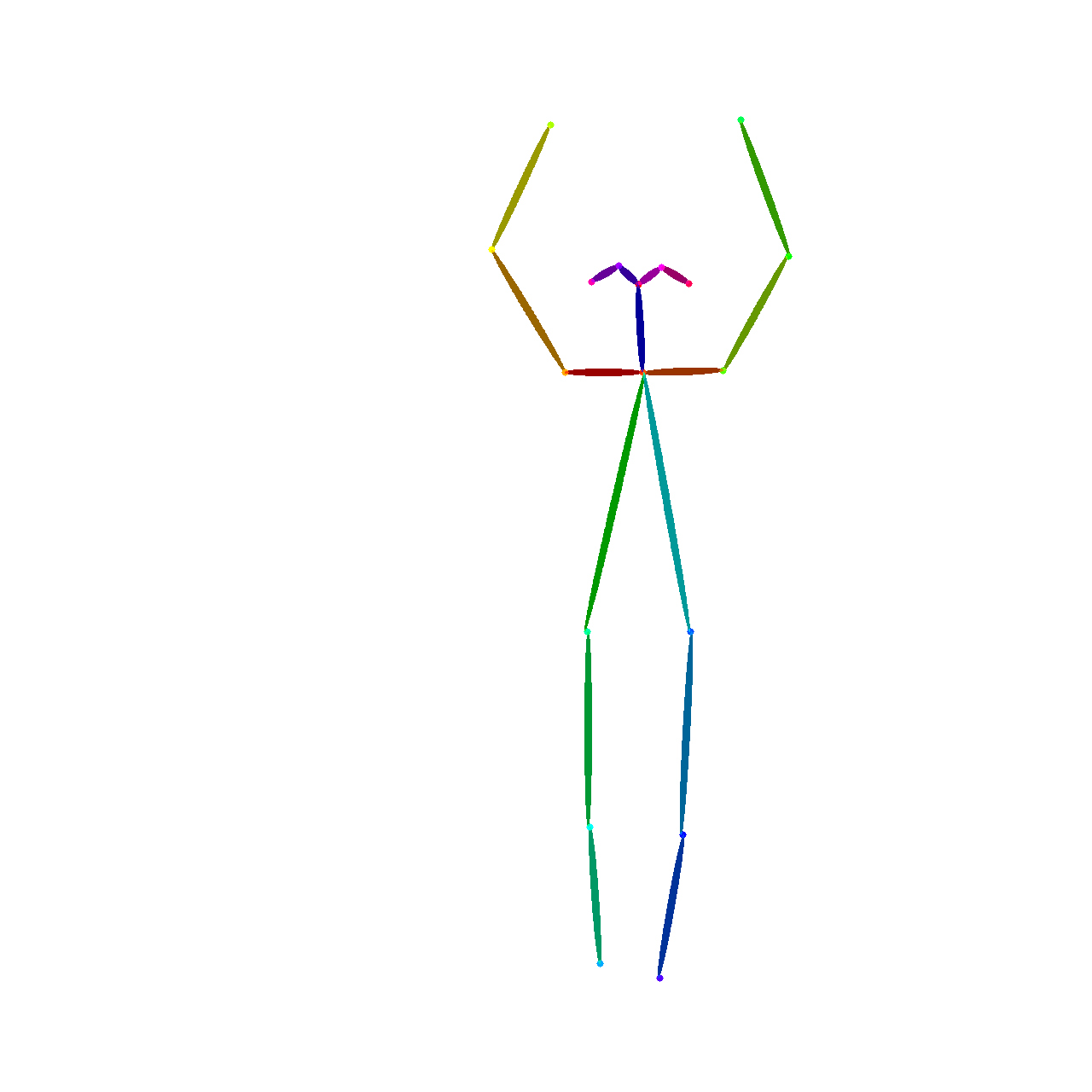}
    \hspace{1mm}
    \imgwrapperNew{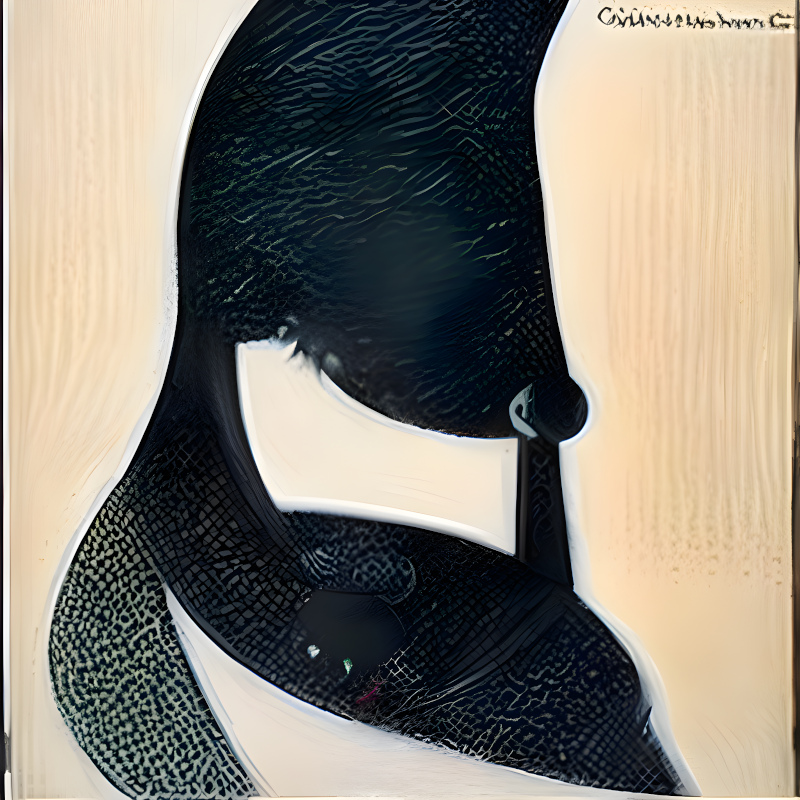}{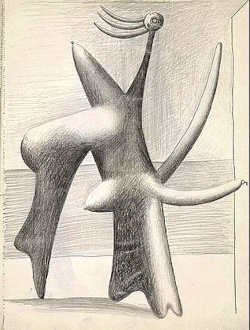}{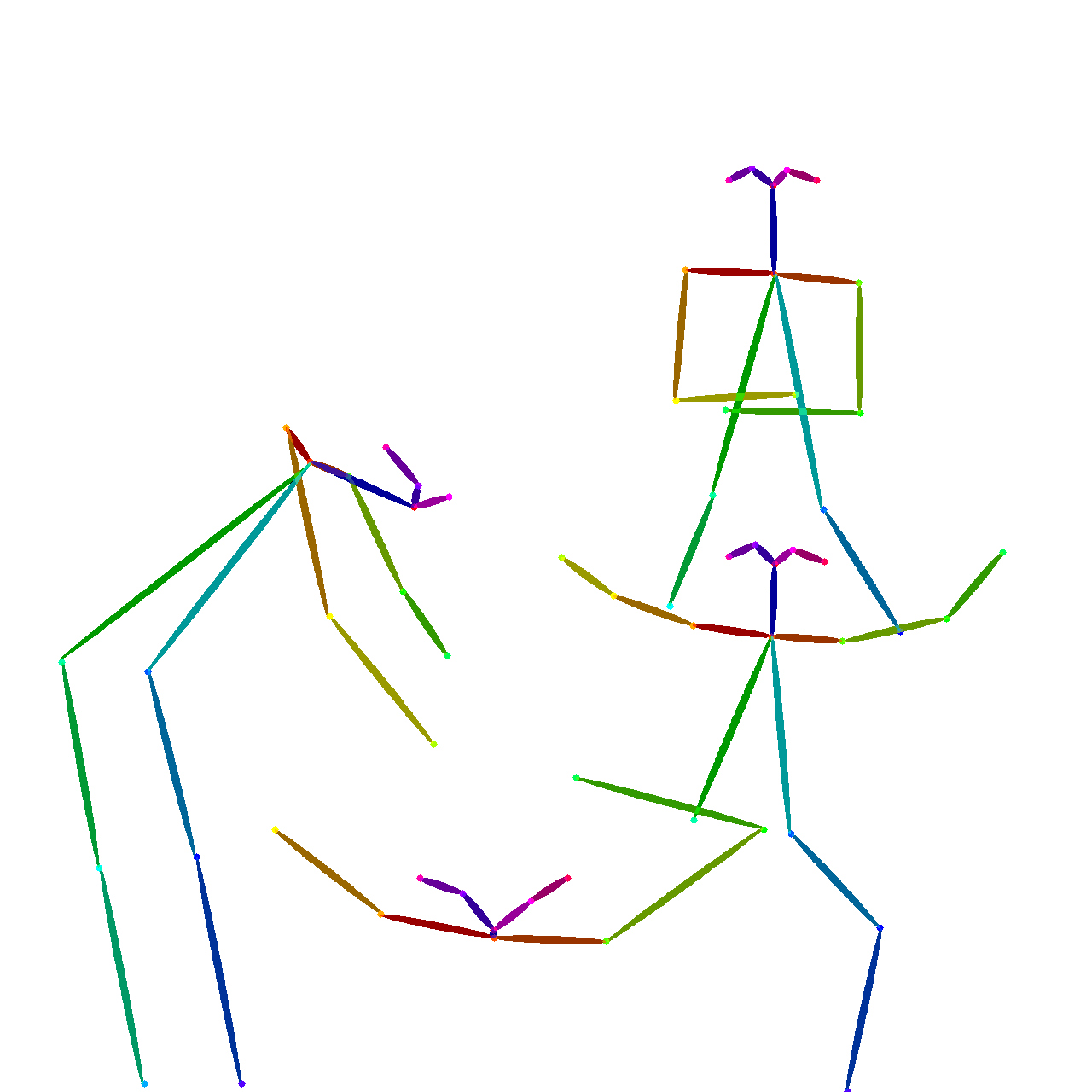}
    \hspace{1mm}
    \imgwrapperNew{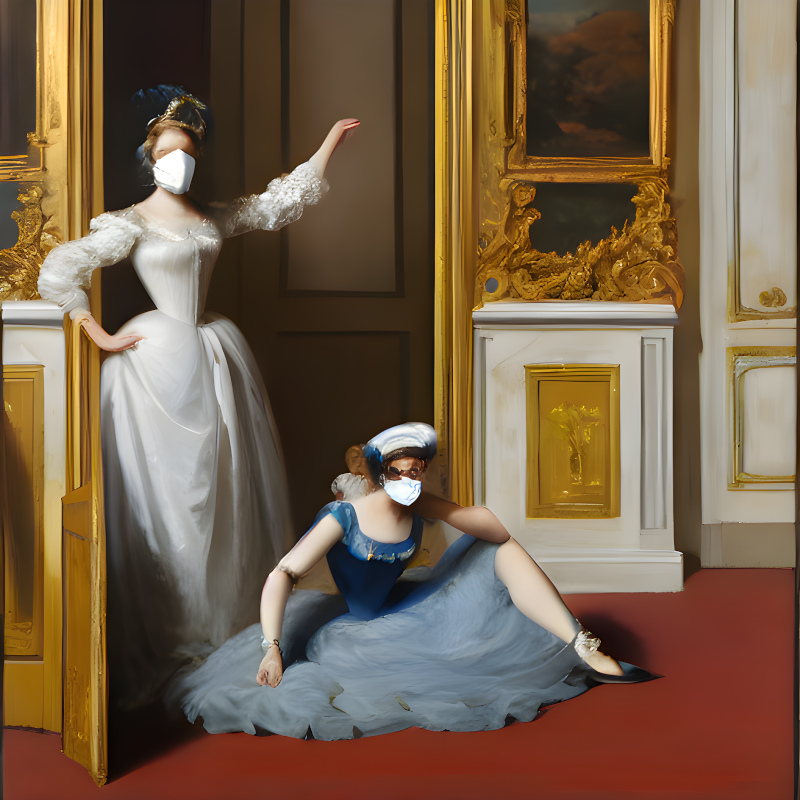}{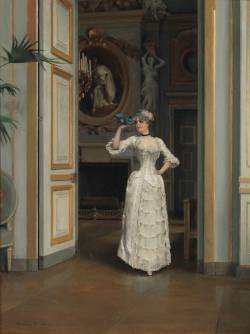}{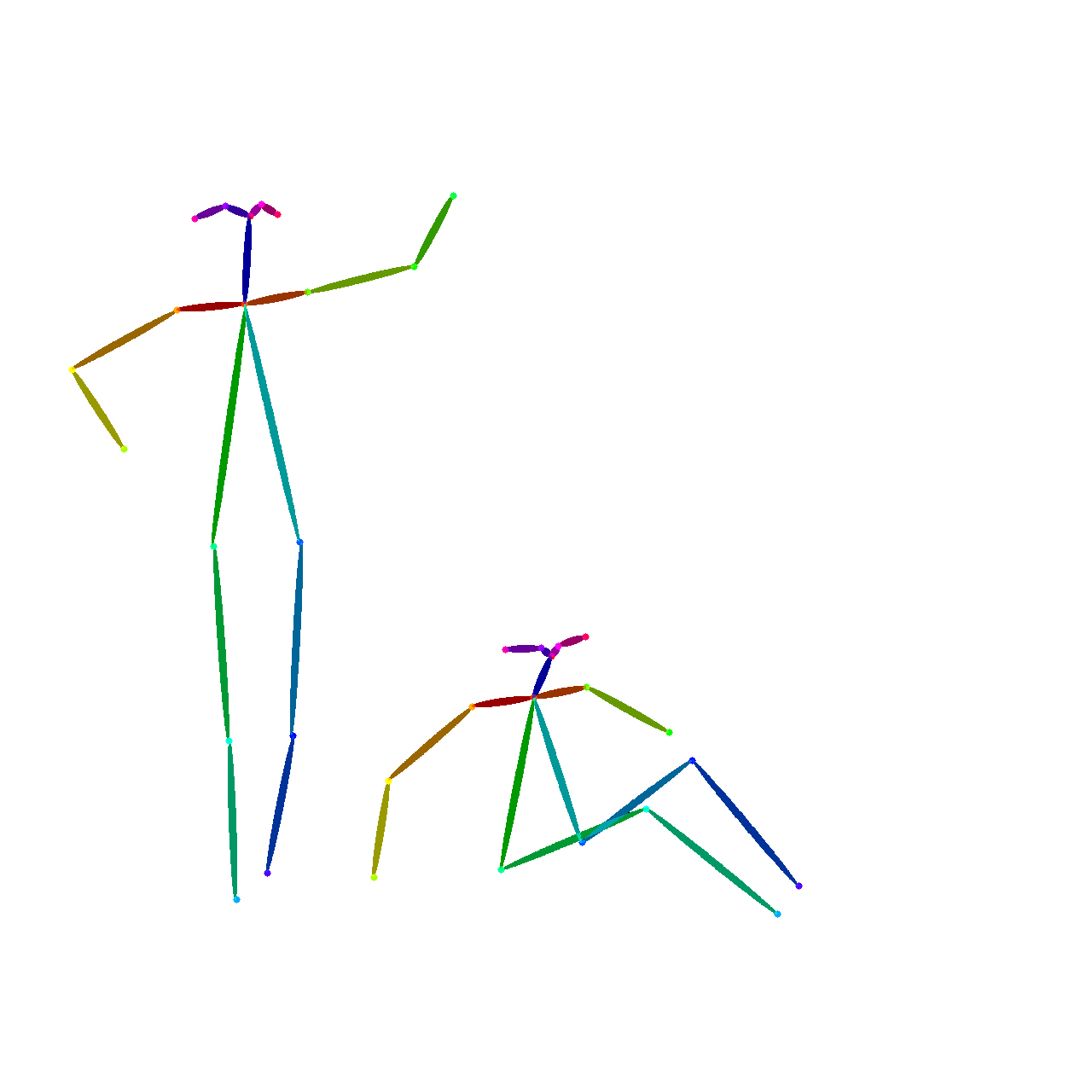}

  \caption{Examples of a change in narrative originating from the user (top row), the generative model (middle row), and both (bottom row), with the respective source artwork (top right) and body prompt (bottom right).
  Hallucinations from the model included, for instance, additional characters in the generated images (middle row, right, and bottom row, left), but also more subtle changes, such as a change in gender of a subject (middle row, center), face masks (bottom row, right), or a man walking an apple as if it were a dog on a leash (middle row, left).
  }
  \Description{Examples of changes in narrative.}%
  \label{fig:examples}
\end{figure*}%

\subsubsection{RQ2: User experience}%
\label{sec:user-experience}%
\ok{%
This part of our investigation looks at how participants experienced body prompting and how they experienced co-creation with the generative AI.%
}%
\paragraph{Experiencing body prompting}%
\label{sec:body-prompt-experience}%
\ok{%
Participants generally found the experience of body prompting highly pleasant ($Mean=4.58$ on a Likert-scale from 1 to 5, $SD=0.62$).
Creating images via body prompting was deemed to be an overall great experience ($Mean=4.58$, $SD=0.65$).
When asked to describe why, participants provided different explanations. The most common reasons were that participants found it fun, interesting, and easy.
    The ease of the interaction via body prompting
    was commented on often.
    The in-app instructions were easy to understand and follow, providing a \textit{``fluent''} (A38, D70) and \textit{``effortless''} (C43) user experience.
}%

\ok{%
In regard to body prompting, the installation provided a novel experience with interesting results.
\textit{``Combining the own posture with an artwork''} (D49) was perceived as entertaining activity and made for an interesting combination (D75).
The physical interaction with the generative AI was perceived as interesting:
\begin{quote}
    \textit{``It's interesting to try 
    what you can do with the AI and just actually physically do something and see it coming out.''} (D76)
\end{quote}
Body prompting was perceived as fun and lighthearted, with
\textit{``not too much pressure participating''} (A35).
Many participants called it a \textit{``fun experiment''} (e.g., C28, G79)
and some participants appreciated that \textit{``you get to be creative and go crazy''} (C28).
Many participants wanted to try the installation a second time and experiment with different body prompts, such as F62 who wanted to \textit{``try a completely different pose than in the artwork''}.
Speaking about Hokusai's The Great Wave off Kanagawa, H30 commented they \textit{``wanted to see if AI transforms us into humans or will it convert us to waves as there were no humans in the original artwork''}.
As depicted in \autoref{fig:examples} (top right), the AI indeed added two humans in this case, although in other cases, it did not.
}%



\ok{%
A small minority of participants voiced some discomfort and concerns about body prompting.
Most of these comments, however, related to the study design and the technical limitations of the installation.
    For instance, three participants commented on the wait time in between prompting and viewing the results which was a fixed part of our study design.
    Some participants commented on the countdown timer.
        On the one hand, F57 thought \textit{``it was quite a long time to stand and look at your own picture''.}
        On the other hand, some participants wanted to have more time to decide on their body prompt. G66, for instance, thought the picture was taken too quickly and \textit{``there was no time to think''.}
Addressing another technical limitation,
    A24 thought it was a fun experience, but voiced disappointment about the facial expression not being captured by the body prompt.
While our privacy-by-design approach was explicitly appreciated by two participants (G61 and B65),
    a few other participants still voiced concerns about 
    being photographed.
    This included comments from E18 and G78 who categorically do not like being photographed. G78 mentioned the picture-taking being \textit{``a little bit distressing''}.
    A2 thought it was daring to perform in public, and
    C43 voiced concerns about being watched: \textit{``When I started to do my pose I started to think is there someone watching. Other than that it was fun to see what is coming up''}.
}%

\ok{%
We could observe that body prompting by a group of people generally tended to be more entertaining, both for the participants performing the body prompting and the on-looking audience.
Groups of participants developed their own dynamics leading to playful interactions with the image generation system.
Selecting the right source artwork often involved short discussions in which participants agreed on their body prompt.
These group negotiations could, obviously, not be observed in single participants. However, some single participants instead asked 
the audience on what body prompt to perform, turning the audience from passive spectators to active participants.
}%

\ok{%
Overall, the novelty of the experience was appreciated by the participants, as exemplified by D47 who said that \textit{``this was a Friday afternoon's fun thing, an experiment, and something new for me''}.
The fact that no specific body prompting instructions were given during the event was seen as positive by many participants who appreciated being able to apply their creativity. 
    For instance, B77 thought it was \textit{``unclear what to expect''} which 
    was \textit{``a good thing for creativity''}. 
Some participants further referred to the creative aspects of body prompting.
    B77 thought it was a creative experience, \textit{``I would have needed help for making the piece of art [with traditional means]. I felt that there was room for creativity''}.
B77 mentioned body prompting required \textit{``quick enough creativity''} in the sense that it \textit{``does not have to be perfect''}.
F62 voiced an interest in exploring emotions with body prompting:
    \textit{``how artificial intelligence modifies -- and how a work of art can be modified according to one's own emotional scale. It would be interesting to use this to express feelings''}.
}
%
%
%
\paragraph{Experiencing AI co-creation via body prompting}%
\label{sec:AI-cocreation}%
\ok{%
In this section, we describe
how participants experienced the co-creative embodied interaction with the generative AI.
Note that this experience 
is, in our study, 
biased by the pre-selected source artwork. Not every combination of source artwork and body prompt worked equally well.
The participants' experiences are, therefore, diverse and difficult to summarize.
Experiences with AI co-creation 
in our study can broadly be divided into whether the generated image met (or exceeded) the par\-ti\-ci\-pants' expectations or not.
Because the latter category is more insightful due to friction in the experience, we place emphasis on the minority of participants in the latter category.
}%

\ok{%
Among the participants who stated their expectations were met ($n=25$),
participants commented the generated image was as they had imagined.
C3, for instance, said the generated imagine \textit{``expresses [the body prompt] pretty well''}.
Some participants found the image better than expected, often commenting on the funniness of the image.
In this participant group, the generated image matched the dynamism of the body prompt. Participants in this group could recognize themselves in the generated image, 
and described the experience as interesting and fun.
}%

\ok{%
For the participants who encountered some friction ($n=54$), the installation was also fun and interesting and overall a very good experience, with some minor limitations.
Participants often commented on the installation's technical limitations, including its inability to reproduce facial expressions and exact hand gestures. This left room for the generative model to add its own interpretation.
    A23 said that \textit{``the first thing that catches [her] eye is those hands, they somehow look a bit unnatural''}.
    In other cases, issues with distorted 
    body parts were even more pronounced, as A35 remarked, \textit{``the fingers look like spaghetti''} and B68 exclaimed \textit{``look at your head, it's like an alien!''}.
A24 also encountered issues with the facial expressions, commenting that \textit{``the face is completely distorted''}.
Some generated images had an uncanniness and strangeness that is often associated with AI-generated images. 
For some participants, the weird generative artifacts made the image \textit{``pleasant to look at but hard to relate''} (A16).
On the other hand, the uncanniness was positive for some participants, such as A24, who, despite the distorted face, thought \textit{``the face is amazing''}, \textit{``better than [we] thought''}, and \textit{``a positive surprise''}.
Many participants had an expectation that the system would produce their likeness.
    C1, for instance, commented \textit{``I expected that the face would resemble my own face''}.
This was even more pronounced in children, as discussed in Section~\ref{sec:children}.
}%

\ok{%
For a few participants, the 
uncanny artifacts shifted the mood of the image, sometimes making it more gloomy, sinister, or eerie 
than anticipated.
    For instance,
    after seeing the unnatural hands, A35 said \textit{``there is something disturbing''} and A33 commented \textit{``it's scarier than we thought''}.
A17 thought there was \textit{``a more serious atmosphere''} in the generated image.
%
%
Some participants said this shift in mood was unexpected, with a few participants commenting that they had wanted the image to be \textit{``more cheerful''} (F62). B67 commented, after finding the outcome \textit{``sad and gloomy''}, that the generated image \textit{``does not reflect what we meant''}.
    A13 also commented that \textit{``it was supposed to be funny, and this is more grim''}.
On the other hand, one participant (A5) thought the generated image had a nice \textit{``Christmas feeling''} that reminded him of a \textit{``Christmas carol''}.
}%


\ok{%
Some participants expressed their surprise at the generated image.
For instance, G79 pointed out a gender change in the subjects, as depicted in \autoref{fig:examples} (middle row, center): \textit{``I assumed there was a man and a woman in the picture, but they both look like men. The environment also surprises.''}
C1 also commented she \textit{``did not expect this outcome''} and C4 said \textit{``it was fun to try but quite surprising''}.
}%

\ok{%
Participants generally found it difficult to understand why the generative model did what it did (even though the body prompt pose was presented to them).
One participant tried to explain the mismatch between their likeness and the generated subject as being a mixture of characteristics of the source artwork and body prompt: \textit{``it has probably taken characteristics of both the original and me, so now it's fun to compare the original to this''} (F74).
Most participants 
could not articulate reasons for deviations in the generated image from their body prompt or the source artwork, and very few tried to verbalize such explanations.
This is partly due to many participants approaching the installation in a casual interaction without strong expectations. Participants also had difficulty imagining what the generated image would look like.
E12 commented on the, in her case, strong 
difference between expectation and outcome. According to her, \textit{``it has very hard to imagine how the outcome would be. That creates very strong expectations''}. Without any reference in mind, she commented that it \textit{``affected the way you see the picture''} and she also mentioned that seeing the generated images from other people gave her a better reference point and an anchor for her own body-prompted image generation.
}%

\subsubsection{RQ3: Participant behavior}%
\label{sec:rq2}%
\label{sec:behavior}%
\ok{%
The participants' key decisions can be summarized as follows:
\begin{itemize}%
    \item 
        \textit{Artwork choice:}
        52 participants (66\%) selected an international artwork from Wikiart
        and 27 (34\%) a local Finnish artwork.
    \item 
        \textit{Staging area choice:}
        46 participants (58\%) used the installation's public posing stage, compared to 33~participants (42\%) who posed in private.
    \item 
        \textit{Participation mode:}
        44 participants (56\%) participated as a group compared to 35 participants (44\%) who body prompt\-ed individually.
\end{itemize}%
In the following, we describe why participants selected a given source artwork which will provide context for understanding their body prompts.%
}%
%
%
\paragraph{Selecting the source artwork}
\ok{%
Overall, participants spent some time on selecting their artwork which is indicative of the thought they put into their selection.
Participants selected the source artwork for a variety of reasons.
}%

\ok{%
Twelve participants mentioned they selected the source artwork because it was familiar to them.
A few participants ($n=6$) were attracted by specific objects in the source artwork. 
Others were attracted to the style of the artwork ($n=4$) and its mood ($n=4$).
C1 mentioned memories of her childhood attracted her to the artwork, when she \textit{``lived by the lake and did laundry''}.
    C10 mentioned being attracted to the painting because it was modern and abstract.
    B61, on the other hand, preferred realistic source artworks over more abstract ones.
}%


\ok{%
A large proportion of the participants ($n=28$) mentioned they selected the source artwork to either contrast or complement  their chosen approach to body prompting. 
    For instance, the number of subjects in the artwork was often ($n=13$) chosen to match the participants' group size.
    For a few participants, the number of subjects in the artworks 
    was perceived as a limiting factor.
        A24, for instance, selected an artwork with three subjects to match their group size, because : \textit{``it was the only one with three characters''.}
}%

\ok{%
Some participants mentioned choosing an artwork with
a simple pose, as it would be \textit{``easier''} to mimic.
On the other hand, in some cases, the source artwork became the source of inspiration for the body prompt.
    A23, for instance, mentioned the artwork  \textit{``immediately gave [her] an idea as to how [the body prompt] could be implemented''} and A7 stated that
    \textit{``maybe the artwork influenced us rather than the other way around''}.
}%

\transition{%
As mentioned above participants often selected their source artwork prompt to match their posing 
strategy, aiming to either contrast (reimagine) or complement (imitate) the source artwork.
In the following section, we describe the different types of posing strategies.
}%


\paragraph{Deciding on a body prompting 
strategy}
\ok{%
When asked ``Why did you pose the way you did?,'' three distinct body prompting 
strategies emerged.
%
About 30\% of participants body prompted to \textbf{re-create the source artwork} in an attempt to
\textit{``mimic''} (G79), \textit{``imitate''} (G64), \textit{``resemble''} (H11), \textit{``mirror''} (F74), and \textit{``fit the artwork''} (C3).
Other participants simply did not want to \textit{``invent anything different''} (A14).
This may reflect a misunderstanding of the technology's purpose, or
simply a natural human behavior to mimic what we see displayed. 
However, in some cases, imitating the pose was not an easy task.
    E12, for instance, said she \textit{``was trying to replicate the [artwork], but the character had a leg completely up, and I did my best to replicate the pose''}.
Other participants also responded that their body prompt was inspired by the source artwork, such as  C36 who \textit{``felt like [the pose] came from the artwork''}.
A13 were more specific in their imitation of the artwork: \textit{``we saw the artwork had a fishing rod. We thought lets do the same fishing rod. We were just imitating whatever was there in the artwork''.}
A34 responded being motivated by seeking harmony with the source artwork, \textit{``because it would adapt well to it, as it is a similar pose''}.
This was also reflected in E15, who said: \textit{``Well, it had to be just like that grandma pose''}.
B54 also mentioned being \textit{``most inspired by the artwork itself''} and the pose \textit{``felt natural''}.
}%

\ok{%
Among the participants who imitated the source artwork, one motivation was to match its mood and atmosphere.
    C36, for instance, thought the artwork was \textit{``energetic and had some `Let's go' feeling''} and this energy inspired her pose.
B55 was touched by the smile in the source artwork and wanted to reciprocate it.
F62 was also moved by the source artwork and its atmosphere, which inspired her body prompt and \textit{``kind of took [her] along''}.
}%

\ok{%
The second posing 
strategy among another third of the participants was to \textbf{reinterpret the artwork}.
%
Participants in this group often reported wanting to \textit{``contrast''} the source artwork to create a novel image that would significantly diverge from the source artwork.
    A5, for instance, mentioned seeking \textit{``a contrast between the posing and the artwork''} to force a change in narrative and create a novel artwork.
    A33 reported wanting \textit{``to recreate the painting in a new way''}.
D75 mentioned wanting \textit{``to create a different atmosphere''} which hints at the underlying motivation to cause a change in narrative between the original artwork and the generated image.
}%

\ok{%
Experimentation was one motivator in this group, as participants were curious to see whether the installation could reproduce a diverging body prompt and what the reimagined artwork would look like.
Another reason for the 
body prompt was to produce a funny or weird picture. This group had some pragmatic participants who assumed their pose \textit{``because nobody stopped us from doing so''} (G78).
Another motivation among a few participants in this group was to make a pose that was easily recognizable.
    B60, for instance, said he assumed the pose so that he could be recognized among multiple people in the group.
}%



\ok{%
Finally, a third of the participants said they assumed the body prompt for \textbf{no specific reason}.
These participants had no deep thoughts about their body prompt and simply went with their intuition in the moment, with the body prompt being \textit{``the first thing that came to mind''} (A23, A24, C3)
or
    body prompted \textit{``impulsively, on the spur of the moment''} (B77).
    F70 mentioned \textit{``just throwing [himself] into the moment''.}

Among the participants in this group, a few responded that the body prompt was their \textit{``normal''} (neutral) pose which is part of their personality (F70).
As above, curiosity was also a key factor, even if they had no strong body prompting strategy. As A34 stated, she \textit{``just wanted to see the end result'',} and \textit{``did not think about the pose so much''.}
    B60 acknowledged that he \textit{``was just interested to see how the system reacts to my pose''.}
    The choice of body prompt 
    was not as important to this group of participants.
The body prompt was sometimes assumed \textit{``in a hurry''} (A38), without deeper reflection, or based on someone else's suggestion.
    C44, for instance, reported that \textit{``someone suggested me to do a crazy pose''}.
This reflects the quick decision 
not being planned or \textit{``premeditated''} (B71) and often improvised in the moment, 
without a specific strategy. 
}%

\paragraph{Deciding on a specific body prompt and teamwork}%



\label{sec:teaming}%
\ok{%
Participants in groups often collaborated 
by discussing and negotiating which image to select, which strategy to follow, and which body prompt to perform.
The body prompt itself also often involved collaboration. For instance, two participants body-prompted back-to-back in a pose of strength and power (see \autoref{fig:examples}). Another pair of participants collaborated by forming a heart with their arms (see \autoref{fig:artworks}).
In some cases, communication was also observed between the 
performer
and the audience 
who made suggestions on how to
act.
This happened more often when participants posed alone than in a group.
This communication highlights the collaborative aspects of the installation and the importance of meta-communication in the co-creative process \citep{014-iccc20.pdf}.
}%

\ok{%
When asked what 
they wanted to express with their body prompt, participant answers were extremely diverse.
Expressing affection was mentioned by a few participants who wanted to hug others (H39) or portray \textit{``friendship''} (A40) and \textit{``family''} (A41).
Participants had fun at the event which was also reflected in their body prompts.
    B59, for instance, wanted to show that they were \textit{``having fun''},
    and B73 mentioned she \textit{``wanted to make something silly''.}
For some, there was a acting 
involved in their body prompt
as they assumed a pose (presumably) different from their own personality.
    A45, for instance, was inspired by the source artwork and wanted to express \textit{``elegance and athleticism''} with a pose describing \textit{``big movements''}.
    A35 aimed to pose as an \textit{``authoritarian evil power''} trying to \textit{``scare people''}.
    This speaks to the entertaining character and playfulness of the installation.%
}%
%
%
%
\subsubsection{RQ4: Body prompting in public}%
\label{sec:bigfive}%
\label{sec:private}%
\ok{%
Our installation was designed to provide 
a choice of body prompting in a public or private setting.
Almost half of the participants ($n=35$) had no strong opinion about this choice 
and did not reveal 
concerns about privacy.
%
%
A significant number of participants ($n=29$) selected to body prompt in public or private for the pragmatic reason of
having a shorter wait in line.
    H11, for instance, selected the public booth \textit{``because it was free and the other one was occupied''} while
    group A35 
\textit{``wanted the others to see what [they were] doing''} (A35).
Seeing the artwork on the big public screen, as opposed to the smaller laptop screen, also motivated some participants to select the public booth.
}%




\ok{%
A minority 
($n=11$, 14\%) mentioned they selected the private booth out of shyness and not wanting to pose in public
considering that it provided a safe space where they \textit{``could be more [themselves] or do crazier things''} (D76) and there was not \textit{``everyone staring''} (G66, B73), which was perceived as embarrassing by a few participants (e.g., D51, F62).
The private booth provided an environment to this group of participants with \textit{``no rush''} (B72) and \textit{``not so much pressure''} (B71, F53).
    However, F62 called her preference for the private booth \textit{``unnecessary shyness''}, as seeing others perform in public was fun and enjoyable to her, but concerns kept creeping up when she was faced with the decision to pose in public herself.
}%

\ok{%
The Big Five personality factors were found to have no significant correlation with the choice of posing area (public vs private), participation mode (group vs alone), source artwork (international vs local), dynamism of the source artwork, and expressiveness of the body prompt (Spearman’s rank correlation, $p>0.05$), with two significant, but weak exceptions.
%
The first 
correlation was found to be in the personality trait of agreeableness: participants who were more agreeable tended to select a more dynamic source artwork; $\rho=0.237$, $p<0.05$, Cohen $d=0.50$, 95\% CI = [0.04, 1.01].
Second, there was a reverse correlation between extraversion and the expressiveness (dynamism) of the body prompt; $\rho=-0.236$, $p<0.05$, Cohen $d=-0.54$, 95\% CI = [-1.01, -0.10].
This indicates that for participants with high level of extraversion, the expressiveness of their body prompt tended to be low. The finding may appear counterintuitive since extraversion is typically associated with sociability, enthusiasm, and positive emotions, traits that 
may be associated 
with more dynamic and expressive poses. The inverse correlation could suggest that extraverted individuals, perhaps due to their comfort with social interactions, may not feel the need to convey their personality as strongly through a single body prompt, or they may approach body prompting with a different mindset compared to more introverted individuals.
However, the correlation was very weak, as the expressiveness of the body prompt is influenced by multiple factors beyond a participant's level of extraversion, such as the context, the individual's mood in-the-moment, or the cultural background.
}%

\subsubsection{Other observations}%
\label{sec:observations}%
%
%
\ok{%
\paragraph{Key moments during the deployment}%
\label{sec:specificinstances}%
A few exceptional body prompts were performed during the event. 
%
One notable instance was a re-enactment of a marriage proposal 
in the public posing area (see \autoref{fig:artworks}, center)
which 
was a heart-warming moment.
}%

\ok{%
While the majority of generated images accurately represented the participant's body prompt, in a few cases, image generation seemingly failed to reproduce the pose.
An example of this is depicted in \autoref{fig:examples} (bottom row, middle) and \autoref{tab:codebook} where a group consisting of four participants performed a highly complex body prompt in the private booth. It featured one person on the shoulders of another  and one person hanging in between two others.
Unfortunately, while the generative model picked up that there was a tree-like structure with light and shadows in the source artwork,  the resulting image (depicted in \autoref{fig:examples}, bottom center) did not reproduce the group's intricate body prompt. This is likely due to the source artwork being abstract and containing no recognizable human pose.
}%




\paragraph{Age appropriateness of the installation}
\label{sec:children}
\ok{%
The Researchers' Night is an event 
visited by many families with children.
Although children were not part of our interview study,
they were allowed to use the installation in the presence of adult caretakers.
This allowed 
an important observation:
some children became very upset after using the installation.
Two young children started crying after seeing their generated image because they expected to see their likeness in the generated image. 
Clearly, the generative system had an emotional impact on some children, conjuring up negative emotions (see also \citet{Alessandro26032024}).
This happened even when the selected source artwork was 
abstract, as in the case of Jean-Michel Basquiat's artworks.
Clearly, the children had different 
expectations than adults and 
a different appreciation for the generated image. For adults, the reimagined abstract artwork was interesting, but young children were often disappointed because they could not recognize 
themselves in the generated image.%
}%
%
%
%
%
\section{Discussion}%
\label{sec:discussion}%
\ok{%
We evaluated body prompting as a novel interaction mechanism for generative AI art installations in a public event setting.
%
Overall, body prompting is a viable way of interacting with image generation AI,
    providing a highly engaging and fun experience for both the active users and on-looking audience.
Our study identified three body prompting 
strategies employed by participants, focused on 
imitating the source artwork, re-imagining, 
or body prompting by ``just being themselves''.
%
%
%
Personality traits did not play an 
important role in the participants' decisions and body prompts and
the installation provided a novel experience which was appreciated by both extraverts and introverts.
}%


\subsection{Human-AI Co-Creation via Body Prompting}%
\ok{%
Generative AI has ushered in an 
era of human-AI co-creation \citep{062-iccc20.pdf,3519026.pdf,3544548.3580959.pdf}.
The human agency in this co-creative process varies on a spectrum \citep{sheridan.pdf}, from full human autonomy to full machine automation.
In our study, participants experienced human-AI co-creation along this spectrum.
Depending on the seleced source artwork and body prompt strategy, the generative AI had varying levels of influence on the generated image.
In combination with the participants' body prompting strategy, 
the participants' experiences included either re-creation or reinterpretation, as discussed in Section \ref{sec:recreating-art}.
The participants pursued their body prompting strategy with varying levels of tenacity, but most interactions 
were casual, 
as participants picked a level of interaction suitable for their desired engagement~\citep{2702613.2702625.pdf,2470654.2481307.pdf}.
}%

\ok{%
While it can be argued whether the body-prompted images are art \citep{978-3-031-29956-8_13.pdf,Coeckelbergh}, the poietic act of performing a body prompt changed the dynamic of the creative process of image generation, placing the human in the center 
of the co-creative process in a public setting.
In \citeauthor{Kun2024}'s GenFrame, participants turned knobs to control image generation parameters, but otherwise relegated the creative process fully to the machine \citep{Kun2024}.
In our co-creative study, the AI became ``more than just a machine'', with AI and humans being co-creators and co-performers \citep{Coeckelbergh}.
This highlights the importance of interaction design in the design of co-creative systems \citep{3519026.pdf}.
%
}%

\subsection{Ethical Considerations}%
\label{sec:ethical}%
\ok{%
In designing and conducting our study, we considered the ethical implications and potential harms that may emerge due to taking pictures of participants in public.
This section elucidates the steps taken to mitigate potential harm.%
}%
\subsubsection{Consent and data privacy}%
\ok{%
Public photography involves capturing people's visual identities, potentially infringing on their privacy rights. While it is legally permissible in many places to photograph individuals in public places without consent, this can still raise concerns, especially when individuals are the central focus of the photograph.
Although public settings do not guarantee privacy, individuals expect not to be intentionally photographed for studies without their consent. To respect the participants' privacy, we implemented anonymization protocols by storing only the detected pose, not the raw photograph. Participants were informed on the consent screen about data collection, the study's purpose, the non-storage of original photos, and their right to withdraw at any time without penalty.
Further, photographic data, especially when linked to identities, are sensitive in nature. We ensured the use of robust security measures to protect the data from unauthorized access, and we committed to using the collected data solely for the purposes of this study.
}%

\subsubsection{Body shaming and self-esteem}%
\ok{%
Photographic representations can sometimes lead to harmful body comparisons or negative self-perception, especially if individuals are self-conscious about their appearance. We strived to conduct our study in a manner that respects the dignity, privacy, and emotional well-being of our participants.
We took particular care to avoid any forms of body shaming or implicit biases in our study design and data interpretation.
To this end, we only collected the pose of participants (detected with openpose), rather than the original photo.
The presentation of results was, therefore, neutral, focusing on behavioral insights rather than personal attributes. The pose information was coarse, without identifying facial expressions or details.%
}%
\subsubsection{Emotional and psychological impact}
\ok{%
We acknowledge that viewing one's own image unexpectedly, as was the case in aour installation where the user saw a webcam video stream during the countdown, might result in unexpected emotional or psychological reactions in some users.
Further, to minimize potential distress, we implemented a debriefing process where participants were given the opportunity to discuss their feelings about the photographs and the study in general.
This process also served as a platform to offer support and resources, should participants feel negatively impacted.
}%


\subsection{Recommendations for the Design of
Body Prompting Installations
}%
\label{sec:lessons}%
\label{sec:advice}%
\ok{%
In this section, we reflect on the few shortcomings in the design of our study and installation, and draw on this experience and the interviews with participants to provide recommendations for practitioners interested in the design of installations and for researchers conducting user studies involving interactive image generation 
in the context of GLAM institutions or public event settings.%
}%

\paragraph{Multi-display setup and user experience design}%
\ok{%
A first recommendation concerns the setup with two separate displays, one for input and one for output.
The main reason for this split was that we wanted to incorporate the interview procedure into the image generation process to bridge
wait time.
However,
at times,
the interviews became the bottleneck of the installation and participants started queuing to see the results.
While this was not a significant issue at our event where visitors also had to queue at other exhibits, it should be avoided in future studies.
%
\\
\noindent
\textit{Design recommendation:}
Future installations should provide immediate feedback to participants without forced integration of a user study.
Bottlenecks in the user flow should be avoided. 
We 
recommend
showing the resulting images on the same screen as soon as they are available.
If there is a choice between private and public body prompting, this 
needs to be explained at the event, for example with signage.
}%

\paragraph{Manage expectations}%
\ok{%
As mentioned in Section \ref{sec:children},
some participants experienced frustration. 
This was particular pronounced with young children who were often disappointed and did not share their parents' appreciation for the AI-generated images.
This raises a point related to managing expectations: an ``AI Camera Booth'' may raise expectations that the resulting images will closely resemble the body prompting person.
Our study's purpose was slightly different from a ``camera booth'', and we carefully designed our communication materials to avoid setting this expectation.
\\
\noindent
\textit{Design recommendation:} 
The purpose of the installation should be made clear to its visitors. 
Communication materials 
should be carefully designed to set the right expectations.
If what is being provided is advertised as a ``camera booth'', the technology should be able to reproduce the likeness of its users.
If the focus is on reinterpreting (or reimagining) existing artworks, the term camera booth should be avoided.
Facial expressions and hand gestures should be included in the body prompt.
}%

\paragraph{Enable expressiveness}
\ok{%
A body prompt can only carry a limited amount of information.
In our case, 
    some participants reported that the generated image was
    more serious and 
    sinister than expected.
This could be alleviated by including the facial expression in the body prompt and providing more fine-grained control over the image generation.
\\
\noindent
\textit{Design recommendation:} 
Include controls for users to set the mood of the generated image in addition to the body prompt,
for instance with \citet{Kun2024}'s rotating knobs or other means, such as a mood selectors or sliders in the application.
}%

\paragraph{Enhancing user experience with AI co-creativity}
\ok{%
Hallucinations add a surprising and playful element to the co-creative experience. Participants who encountered and noticed these hallucinations were entertained and intrigued.
Uncontrollable AI \citep{8c37b61b693c475096f73bf46a3323df} adds value to the user experience and may foster creativity.
However, the trade-off between user and system agency should be carefully balanced~\citep{ICCC24_paper_15.pdf}.
\\
\noindent
\textit{Design recommendation:} 
Hallucinations and weird artifacts are part of the generative AI user experience and add an intriguing component.
We recommend not seeking to eliminate hallucinations entirely. Instead, the the system should embrace hallucinations and surprising interactions in a playful way.
}%

\paragraph{Interaction with the installation}
\ok{%
In our installation, in order to body prompt, users first needed to select a source artwork.
While this interaction could take place with gestures, we decided on a touch-screen display. However, users may be unaware of interactive content \citep{henderson2017.pdf}.
We observed this interaction blindness in one case at the event where participants in the private booth faced the ``first-click problem'' -- they 
did not realize the display was a touchscreen and used a mouse and keyboard that happened to be available.
With the public installation, this was not an issue since participants could observe other 
participants' use of the installation.
\\
\noindent
\textit{Design recommendation:} 
The first-click problem should be addressed with instructions that are carefully designed to fit the study purpose and not bias participants.
While we addressed this problem with a makeshift banner during the event, the private booth requires a slightly different design and instructions than the public posing stage.
}%

\paragraph{Takeaways and souvenirs}
\ok{%
Many participants liked the resulting images and inquired about ways of downloading them. These participants developed a sense of ownership of the generated images.
In our event, we strived for a minimal setup and could not facilitate the download of images. Instead, participants were told to take an image of the artwork with their mobile phone.
\\
\noindent
\textit{Design recommendation:}
Provide a means for visitors to take their body prompted image home with them. This could be in the form of a technical solution, such as a QR code for downloading the generated images, or via traditional means, such as print-outs.
}%

\paragraph{Nudity and inappropriate imagery}%
\ok{%
Stable Diffusion has been shown to generate unsafe images and hateful memes \citep{2305.13873.pdf}.
Our own early testing during the development of the installation confirmed that generated images can contain accidental nudity and unsafe imagery, if not safeguarded.
\\
\noindent
\textit{Design recommendation:} 
Guardrails should be implemented to prevent unsafe generations in public installations.
We mitigated this issue 
by
carefully selecting source artworks, avoiding nudity and blood, 
and using a ``negative'' prompt designed to prevent the image generation model from producing explicit imagery.
The design of this prompt was based on best practices used in the community of text-to-image generation practitioners.
}%

\paragraph{Materials}
\ok{%
The selection of artworks with a popular, but also local context proved to be a good design choice, as the local art was well received.
However, in some cases, there were combinations of body prompts and source artworks that did not work well together.
\\
\noindent
\textit{Design recommendation:}
Provide a variety of source artworks for visitors to experiment with,  but ensure that they are related to the human figure.
In particular, provide source images that feature different 
subjects and compositions to cater to the body prompting 
strategy of imitation.
Provide source artworks with different levels of complexity and body prompting difficulty, leaving the choice 
to the user.
At the same time, in order to allow for a more consistent comparison of the AI-generated images, it is critical to establish more specific criteria for the selection of the source artworks, for e.g. prioritizing figurative painting to the detriment of abstract art.
}%

\paragraph{Personality traits}
\ok{%
Personality traits, such as extraversion and openness, were not a significant factor in our study.
This is, in part, due to the specific context of the event that attracts a self-selected sample of participants with tendencies to being extraverted. The Researchers' Night is visited by people curious to make new experiences, and our installation successfully provided these experiences.
In this regard, the value of having a separate private posing booth is questionable.
A related issue is the nature of the private booth: it is hidden from view and easy to overlook. However, even if the booth had been better advertised, the general tendency among participants was that public booth was the preferred option, which is reflected in the visitor numbers and participant responses.
\\
\noindent
\textit{Design recommendation:} 
We recommend not creating a private booth, and instead focus all efforts on designing the public posing stage for effective engagement with the audience.%
}%

%
\section{Conclusion}%
\label{sec:conclusion}%
%
%
\ok{%
Generative AI opens new opportunities and alternative ways to engage 
audiences in image generation with generative AI.
Embodied interaction for co-creating images with generative AI is an intriguing possibility, providing a novel, fun, and highly engaging experience.
Our study explored 
body prompting
as a way to interact with image generation AI via an art installation  in a public event setting.
As demonstrated by our study, such interaction acts as a meaningful connector between historical artistic practices, such as the 
act of posing for a portrait, and new creative processes involving AI technology.
As observed in our installation, generative AI provides a means for effectively fostering personal expression and collaborative co-creation.
Our recommendations serve as guidance for future studies and the design of installations involving generative AI in the context of GLAM institutions and other public settings.
}%
%
%
%
%
    \section*{Acknowledgements}
    This work was partially supported by the Natural Sciences and Engineering Research Council of Canada (NSERC).
    We thank reviewers for their valuable input.
    We are grateful to the research assistants Aleksandra Hämäläinen, Saila Karjalainen, and Petri Salo for their 
    contributions to this research project. 
\bibliographystyle{ACM-Reference-Format}
\bibliography{paper}

\appendix
\section{Codebook}
\label{appendix:codebook}
\newcolumntype{Y}{>{\raggedright\arraybackslash}X}
\begin{table*}[htb]%
\caption{Codebook for coding the expressiveness (dynamism) of the participant's body prompt and source artwork.}%
\Description{}
\label{tab:codebook}%
\footnotesize%
\begin{tabularx}{\textwidth}{X|YY|YY|YY}%
\toprule%
    & \multicolumn{2}{c}{low}  & \multicolumn{2}{c}{medium}  & \multicolumn{2}{c}{high}  \\
\midrule%
    Dynamism of \newline source artwork &
    static subjects; straight lines or absence of diagonal lines   &
\parbox[t]{\linewidth}{%
\raisebox{-.9\height}{%
\includegraphics[width=\linewidth]{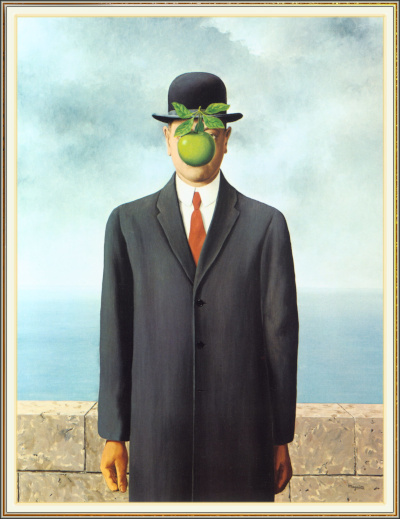}%
}%
}%
&
    some motion; some diagonal lines &
\parbox[t]{\linewidth}{%
\raisebox{-.85\height}{%
\includegraphics[width=\linewidth]{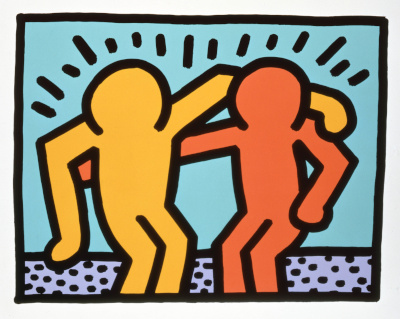}%
}%
}%
&
    presence of diagonal lines; demonstrating strong motion; subject positioned along diagonal axis &
\parbox[t]{\linewidth}{%
\raisebox{-.9\height}{%
\includegraphics[width=\linewidth]{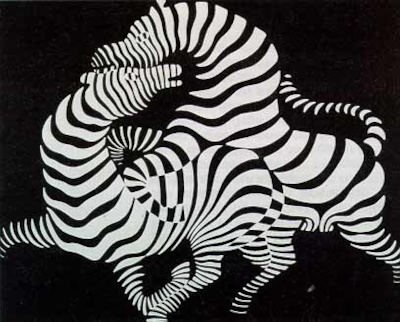}%
}%
}%
    \\
\midrule
    Dynamism of \newline body prompt &
    static pose; standing relaxed; no raised limbs &
\parbox[t]{\linewidth}{%
\raisebox{-.9\height}{%
\includegraphics[width=\linewidth]{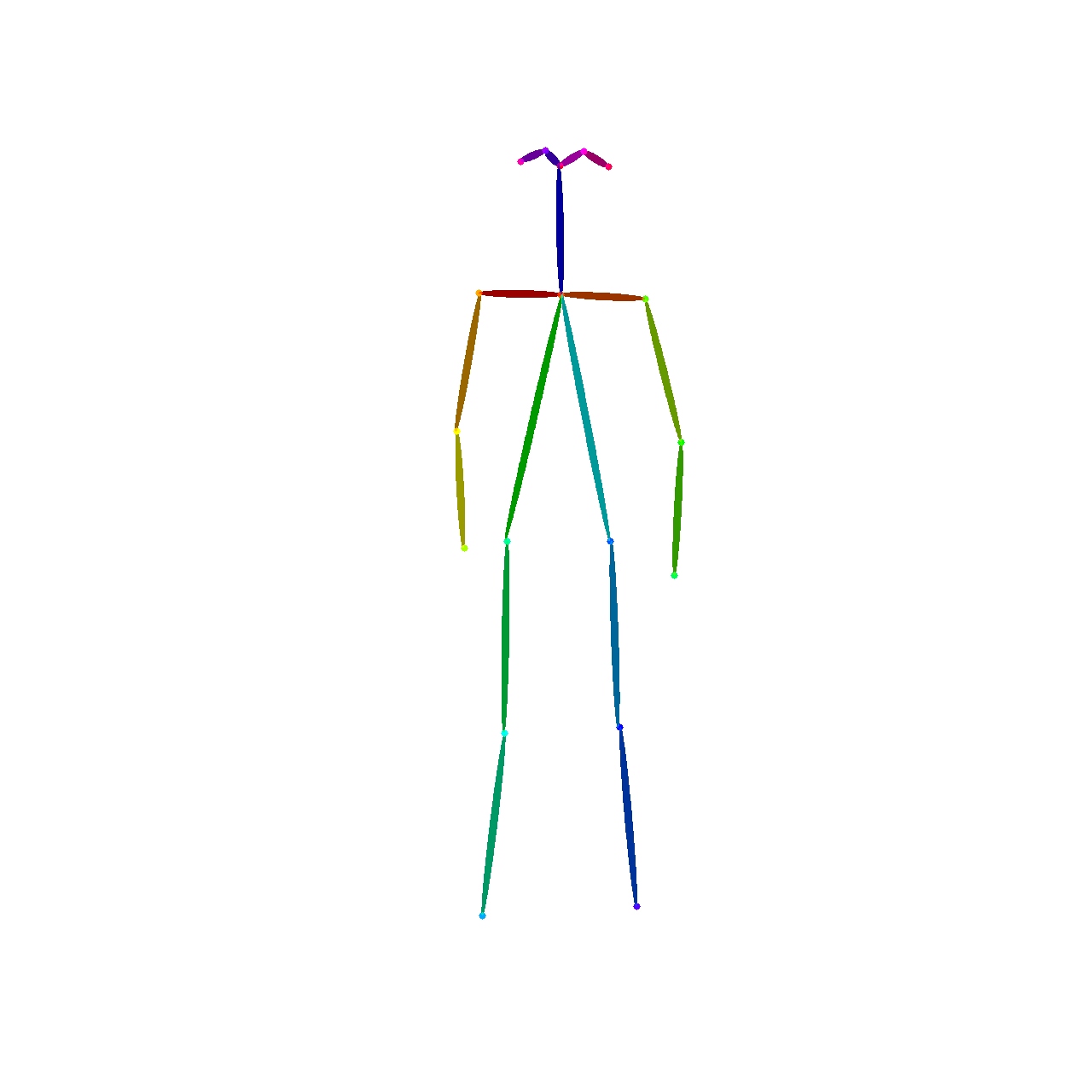}
}%
}%
&
    some expression; one raised limb or a strong static pose (other than standing relaxed) &
\parbox[t]{\linewidth}{%
\raisebox{-.9\height}{%
\includegraphics[width=\linewidth]{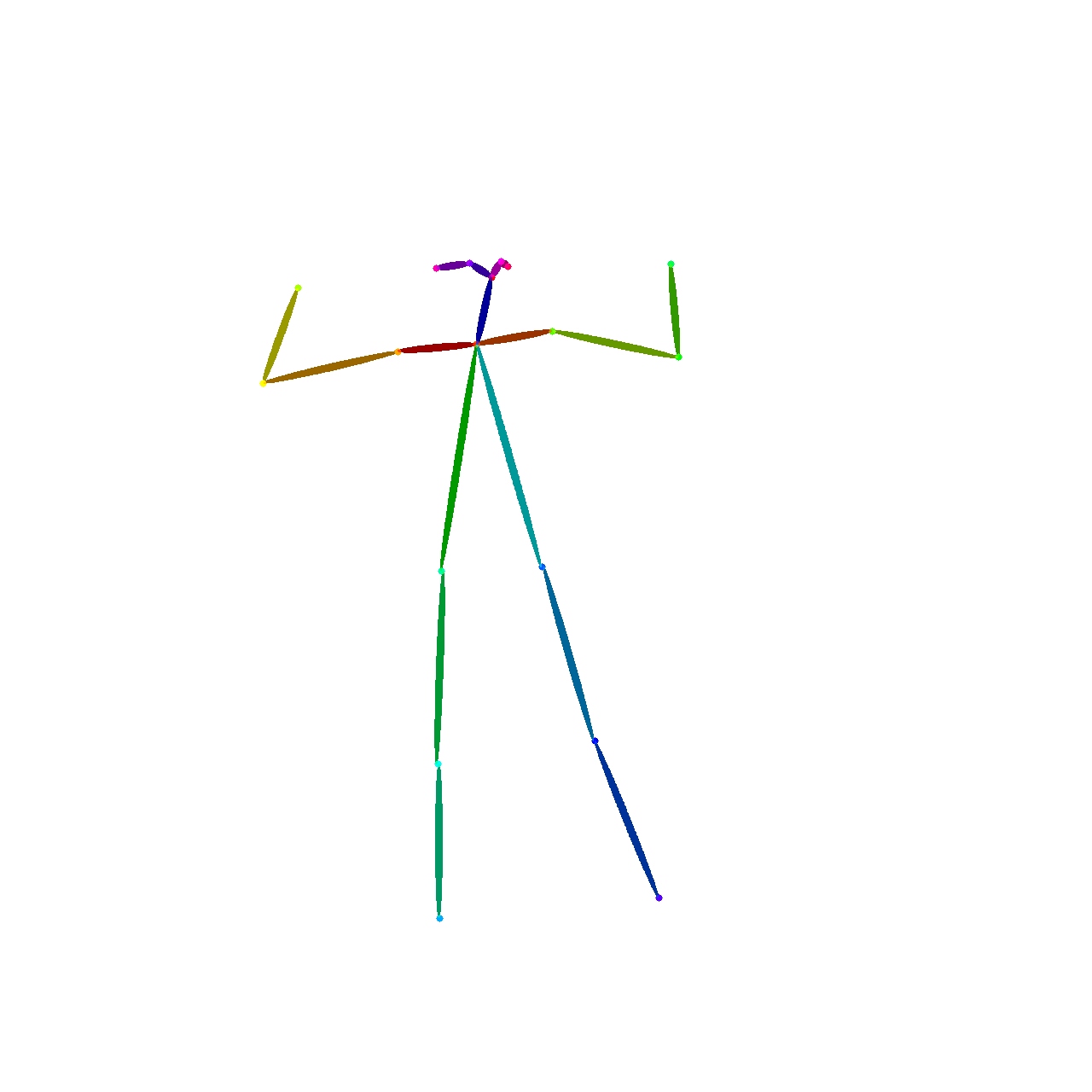}
}%
}%
&
    strongly expressive pose; raised legs; more than just arms involved &
\parbox[t]{\linewidth}{%
\raisebox{-.9\height}{%
\includegraphics[width=\linewidth]{figures-JPG/artworks/private/1696010254.0216322/openpose-transparent.jpg}%
}%
}%
    \\
\bottomrule%
\end{tabularx}%
\end{table*}%

\end{document}